\documentstyle[aps,preprint,eqsecnum]{revtex}
\begin{document}
\preprint{WUGRAV-94-6}
\preprint{NU-GR-11}
\draft
\title{Coalescing binary systems of compact objects
to (post)$^{5/2}$-Newtonian order. \\
V. Spin Effects}

\author{Lawrence E. Kidder
\thanks{Present address:  Department of Physics and Astronomy,
Northwestern University, Evanston, Illinois 60208.}}
\address{
McDonnell Center for the Space Sciences, Department of
Physics,
Washington University, \\ St. Louis, Missouri 63130}
\date{\today}
\maketitle

\begin{abstract}

We examine the effects of spin-orbit and spin-spin coupling on the inspiral of
a coalescing binary system of spinning compact objects and on the gravitational
radiation emitted therefrom.  Using a formalism developed by Blanchet, Damour,
and Iyer, we calculate the contributions due to the spins of the bodies to the
symmetric trace-free radiative multipole moments which are used to calculate
the
waveform, energy loss, and angular momentum loss from the inspiralling binary.
Using equations of motion which include terms due to spin-orbit and spin-spin
coupling, we evolve the orbit of a coalescing binary and use the orbit to
calculate the emitted gravitational waveform.  We find the spins of the bodies
affect the waveform in several ways:  1) The spin terms contribute to the
orbital decay of the binary, and thus to the accumulated phase of
the gravitational waveform.  2) The spins cause the orbital plane to precess,
which changes the orientation of the orbital plane with respect to an
observer, thus causing the shape of the waveform to be modulated.  3)
The spins contribute directly to the amplitude of the waveform.
We discuss the size and importance of spin effects for
the case of two coalescing neutron stars, and for
the case of a neutron star orbiting a rapidly
rotating $10M_\odot$ black hole.

\end{abstract}

\pacs{04.25.Nx, 04.30.-w, 04.80.Nn, 97.80.Fk}

\narrowtext

\section{Introduction and summary}

Coalescing binary systems of compact objects
are the most promising source of gravitational waves which could be
detected by laser interferometric gravitational wave detectors such as the
currently funded U.S. LIGO and French/Italian VIRGO detectors \cite{ligo}.
These systems consist of neutron stars or black holes whose orbits decay
because
of the dissipative effect of gravitational radiation.  The binary pulsar
PSR 1913+16 is an example of such a system and has given us our first
evidence that gravitational waves exist
\cite{binary_pulsar}.  Laser interferometric gravitational
wave detectors will be able to observe the very late stages of the inspiral
of coalescing binaries (typically the final several minutes) as the
gravitational wave frequency sweeps through a detector's bandwidth from 10 Hz
to 1000 Hz.

In the previous papers in this series \cite{linc,agw92,isco,agwtail}
we have studied the evolution of coalescing binary systems
using a post-Newtonian approximation.  The post-Newtonian
approximation involves an expansion of
corrections to Newtonian gravitational theory
with an expansion parameter
$\epsilon \approx v^2 \approx m/r$,
which is assumed to be small (we use units in which $G=c=1$), where
$m=m_1 + m_2$ denotes the
total mass of the system, and
where $r$ and $v$ are the orbital separation and velocity.
Since we are interested in the late stages of
the inspiral, where the fields may not be so small, and the velocity not so
slow, we use an expansion carried out to the highest practical order.
We use equations of motion carried out to
(post)$^{5/2}$-Newtonian order,
the order at which the dominant gravitational
radiation-reaction damping forces occur.
Schematically the equations of motion are given as
\begin{eqnarray}
d^2 {\bf x}/dt^2 = -(m{\bf x}/r^3)
[ && 1+{\cal O}(\epsilon)+{\cal O}(\epsilon^{3/2})+{\cal O}({\epsilon}^2)
\nonumber \\ && \mbox{}
 + {\cal O}({\epsilon}^{5/2})+ \cdots ],
\end{eqnarray}
where ${\bf x} = {\bf x_1} - {\bf x_2}$ denotes the separation vector
between the bodies
and $r=\vert \bf x \vert$.
We use a gravitational waveform carried out to
(post)$^{3/2}$-Newtonian order
beyond the quadrupole formula.  Schematically,
\begin{equation}
h^{ij} = {2 \over D} \left[ Q^{ij} \left\{ 1 + {\cal O}(\epsilon^{1/2})
+ {\cal O}(\epsilon)
+ {\cal O}(\epsilon^{3/2}) + \cdots \right\} \right]_{TT} ,
\end{equation}
where $Q^{ij}$ represents the usual quadrupole term (two time derivatives of
the mass quadrupole moment tensor), $D$ is the distance between the source
and an observer, and $TT$ denotes that the
transverse traceless part of the tensor should be taken ((post)$^2$-Newtonian
contributions have recently been derived by Blanchet {\it et al} \cite{2pn}).
Given a set of initial conditions, we evolve the orbit with the equations of
motion and calculate the emitted gravitational radiation.
In previous papers we ignored the effects on the motion and radiation
due to the spins of the bodies.  It is the purpose of this paper to take these
effects into account.

The contribution of the spins of the bodies to the equations of motion has been
studied by numerous authors \cite{barker,hartle,damour300,brumberg}.
They include a contribution due to a spin-orbit
interaction, and a contribution due to a spin-spin interaction.  In Sec.
\ref{sec:orbit} we
write down the spin-orbit and spin-spin contributions to the equations of
motion and show that, although they are formally post-Newtonian corrections
to the equations of motion, for compact bodies they are effectively
(post)$^{3/2}$-Newtonian and (post)$^2$-Newtonian corrections
respectively, in part because they involve factors of
order $(R/r)$ where $R$ is the size
of the body which is on the order of $m$ for a compact body.
We add these contributions to our previous equations
of motion, and obtain equations of motion valid for arbitrary masses and spins.
Our equations of motion neglect tidal effects; for compact
binary systems these effects
are expected to be very small until the very late stages of inspiral
\cite{linc,bildsten}.  We also
ignore rotationally induced quadrupole effects which would enter at the same
order as spin-spin effects; these also are expected to be small until very
late stages, except possibly for very rapidly rotating Kerr black holes
\cite{bildsten}.
The major effect of the spins on the orbital evolution is that they cause
the orbital plane to precess, thus changing its orientation in space.  The
spins themselves also precess; the precession equations are also written down
in
Sec. \ref{sec:orbit}.

Recently, we calculated the spin contributions to the symmetric trace-free
radiative multipole moments which we used to calculate the spin contributions
to
the energy lost from a binary system due to the gravitational radiation
emitted from the system \cite{kwwspin}.
In Sec. \ref{sec:rad} we present the details of that calculation, and calculate
the spin contributions to the gravitational waveform, the angular momentum
lost from the system, and the linear momentum ejected from the system.

Since the emission of gravitational radiation tends to circularize the
orbits of the binary system \cite{linc},
we study the system under the assumption
of circular orbits.  This assumption should hold until the very late stages
of inspiral when either the bodies start to merge, tidal effects become
important, or the innermost stable circular orbit is reached in which case
the system undergoes a transition from inspiral to plunge \cite{isco,kww}.
Assuming circular
orbits, we calculate the spin-orbit and spin-spin contributions to the
energy loss and variation of the orbital
frequency, and hence to corrections in the accumulated orbital phase.
In Table \ref{phasetable} we compare these
contributions to the analogous contributions
due to the quadrupole term and other post-Newtonian terms for various binary
systems.
These contributions to the orbital phase are important since, in order
to extract information from the observed waveforms, theoretical templates
must match the observed waveform to within about one
cycle (which corresponds to
half an orbit) over the several hundred cycles
that appear in the sensitive region of a detector's bandwidth
(between roughly 40 Hz and 100 Hz).  Note that the templates do not have to
match the observed waveform within one cycle over the
entire bandwidth from
10 Hz to 1000 Hz (as suggested by Cutler {\it et al}
\cite{jugger}) since the detector is more sensitive to the signal at some
frequencies than in others \cite{sam}.  As seen in Fig. 2 of Ref.
\cite{flanagan},
60\% of the signal-to-noise from a binary inspiral will accumulate between 40
Hz
and 100 Hz due to the shape of the LIGO noise spectrum, and to the fact that
there
are more orbits at lower frequencies.
In Table \ref{phasetable} we list the orbital phase contributions for both
the restricted bandwidth (40-100 Hz) and the entire bandwidth (10-1000 Hz)
of a LIGO-type detector.
We see that the spin-orbit correction to the accumulated orbital
phase is important (unless the spins are very small), while the
spin-spin contribution is negligible unless rapidly rotating black holes are
present.  Note that the contributions of the post-Newtonian terms are less
important in the narrower bandwidth than in the entire bandwidth since
the sensitive region of the bandwidth is at low frequencies, for which
post-Newtonian corrections are less important in most systems.
Conversely, the use of matched templates may allow estimations of
spins via the spin-orbit terms \cite{kwwspin,jugger}.

Using the equations of motion and the equations of precession, we numerically
evolve the orbit for a binary system of arbitrary masses and spins.  Using
the orbit we are then able to calculate the gravitational waveform emitted by
the system.
If the spins of the bodies are not aligned perpendicular to the
orbital plane, the orbital plane will precess in space thus changing its
orientation with respect to an observer.  Since an observed gravitational
waveform depends upon the orientation of the orbital plane with respect to
the observer, this precession will cause the waveform to be modulated.
Fig. \ref{amplitude_modulations}
shows an example of this modulation.  We see that these modulations
depend significantly upon the observer's location with respect to the source,
and upon the orientation of the detector relative to the incoming wave.
The size of the modulations also depends upon the relative magnitudes
 of the orbital
angular momentum ${\bf L}$ and the total spin ${\bf S}$ and upon their relative
orientation.  (See Fig. \ref{source_coordinates} for a description of
the source coordinate system.)  If the angle
between ${\bf L}$ and ${\bf S}$ is small, the
modulations of the waveform will be small.
If $|{\bf S}| << |{\bf L}|$, then the modulations are small
regardless of their relative orientation.
For a circular orbit
$|{\bf L}| = \mu (rm)^{1/2}$ to leading order, while for each spin we define
$|{\bf S_A}| = \chi_A m_A^2$ where $0 \leq \chi_A \leq \chi_{max}$, where
$\chi_{max} = 1$ for a black hole, and depends on the uncertain nuclear
equation of state for neutron stars.  For most neutron star models
$\chi_{max} \leq 0.7$ \cite{nsspin}.  In Fig. \ref{momentum_size}
we compare the relative
sizes of ${\bf L}$ and ${\bf S}$ for the case of equal masses and for
the case of a 10:1 mass ratio.
We see that
$|{\bf S}| << |{\bf L}|$ almost
always for the equal mass case, so that the modulations of the waveform
will be small.
For a neutron star orbiting a rapidly rotating massive black hole, however, the
modulations can be substantial if ${\bf L}$ and ${\bf S}$ are sufficiently
misaligned.
The modulation of the waveform due to the spin-induced orbital precession
has also been independently studied by Apostolatos {\it et al}
\cite{precession}.

There is also an explicit contribution to the waveform due to radiative
multipole moments generated by the spins of
the bodies.  This contribution is relatively small until the late stages of
inspiral.  In Fig. \ref{gravitational_waveform}
we compare the spin-orbit and spin-spin
contributions to the waveform  with
the quadrupole contribution and higher-order post-Newtonian contributions for
the last few orbits before coalescence.  We see that the spin-orbit
contribution is comparable to the higher-order post-Newtonian contributions,
while the spin-spin contribution is practically negligible.

The rest of the paper presents the details.  In Sec. \ref{sec:orbit}
we assemble the
equations necessary to evolve the orbit.  In Sec. \ref{sec:rad}
we derive the spin
corrections to the radiative multipoles, and calculate the spin corrections
to the gravitational radiation emitted, energy lost, angular momentum lost, and
linear momentum ejected from the binary system.  In Sec. \ref{sec:circ}
we examine the
system in the limit of circular orbits, and in the
limit of small precessions.  In
Sec. \ref{sec:results} we present our results
for various numerically evolved orbits.  Finally, in Sec. \ref{sec:conclusion}
we discuss these results, and their
implications for extracting useful information
from observations.  In Appendix \ref{sec:appA} we
discuss the issue of ``spin supplementary
conditions"  used to fix the center of mass of the spinning bodies to
post-Newtonian order.  In Appendix \ref{sec:appc} we list the post-Newtonian
corrections to the circular orbit waveform.
In Appendix \ref{sec:appB} we compare our results with
calculations involving test masses orbiting spinning black holes.

\section{Orbital Evolution Equations}
\label{sec:orbit}

\subsection{Equations of Motion}

Equations of motion for two bodies of arbitrary mass and spin have been
developed by numerous authors (for reviews and references see
\cite{barker,hartle,damour300,brumberg}).
By eliminating the center of mass of the system, we convert the two body
equations of motion to a relative one-body equation of motion given by
\begin{equation}
{\bf a} = {\bf a}_N + {\bf a}_{PN} + {\bf a}_{SO} + {\bf a}_{2PN}
+ {\bf a}_{SS} + {\bf a}_{RR} , \label{pneom}
\end{equation}
where
\widetext
\begin{mathletters}
\begin{equation}
{\bf a}_N = - {m \over r^2} {\bf \hat n} ,
\end{equation}
\begin{equation}
{\bf a}_{PN}=  - {m \over r^2} \left\{  {\bf \hat n} \left[
(1+3\eta)v^2 - 2(2+\eta){m \over r} - {3 \over 2} \eta \dot r^2 \right]
-2(2-\eta) \dot r {\bf v} \right\} ,
\end{equation}
\begin{equation}
{\bf a}_{SO} = {1 \over r^3} \left\{ 6 {\bf \hat n} [( {\bf \hat n} \times
{\bf v} ) {\bf \cdot} (2{\bf S} + {\delta m \over m}{\bf \Delta} )]
- [ {\bf v} \times (7
{\bf S}+3{\delta m \over m}{\bf \Delta})]
+ 3 \dot r [ {\bf \hat n} \times
(3{\bf S} + {\delta m \over m}{\bf \Delta} )] \right\} ,
\end{equation}
\begin{eqnarray}
{\bf a}_{2PN} = - {m \over r^2} \biggl\{ && {\bf \hat n} \biggl[ {3 \over 4}
(12+29\eta) ( {m \over r} )^2
+ \eta(3-4\eta)v^4 + {15 \over 8} \eta(1-3\eta)\dot r^4
\nonumber \\ && \mbox{}
- {3 \over 2} \eta(3-4\eta)v^2 \dot r^2
- {1 \over 2} \eta(13-4\eta) {m \over r} v^2
- (2+25\eta+2\eta^2) {m \over r} \dot r^2 \biggr]
\nonumber \\ && \mbox{}
- {1 \over 2} \dot r {\bf v}
\left[ \eta(15+4\eta)v^2 - (4+41\eta+8\eta^2)
{m \over r} -3\eta(3+2\eta) \dot r^2 \right] \biggr\} ,
\end{eqnarray}
\begin{equation}
{\bf a}_{SS} = - {3 \over \mu r^4} \biggl\{ {\bf \hat n} ({\bf S_1 \cdot
S_2}) + {\bf S_1} ({\bf \hat n \cdot S_2}) + {\bf S_2} ({\bf \hat n \cdot
S_1}) - 5 {\bf \hat n} ({\bf \hat n \cdot S_1})({\bf \hat n \cdot S_2})
\biggr\} ,
\end{equation}
\begin{equation}
{\bf a}_{RR} = {8 \over 5} \eta {m^2 \over r^3} \left\{ \dot r {\bf \hat n}
\left[ 18v^2 + {2 \over 3} {m \over r} -25 \dot r^2 \right] - {\bf v}
\left[ 6v^2 - 2
{m \over r} -15 \dot r^2 \right] \right\} ,
\end{equation}
where ${\bf a}_N$, ${\bf a}_{PN}$, and ${\bf a}_{2PN}$ are the Newtonian,
(post)$^1$-Newtonian, and (post)$^2$-Newtonian contributions to the
equations of motion, ${\bf a}_{RR}$ is the contribution to the equation
of motion due to the radiation-reaction force, and
${\bf a}_{SO}$ and ${\bf a}_{SS}$ are the spin-orbit and spin-spin
contributions to the equations of motion which we have ignored previously,
and where ${\bf x} \equiv {\bf x_1}-{\bf x_2}$, ${\bf v}={d{\bf x}/dt}$,
${\bf \hat n}\equiv{{\bf x}/r}$, $\mu \equiv m_1m_2/m$, $\eta \equiv
\mu /m$, $\delta m \equiv m_1 - m_2$, ${\bf S} \equiv {\bf S_1}+{\bf S_2}$,
and ${\bf \Delta} \equiv m({\bf S_2}/m_2 -{\bf S_1}/m_1)$, and an overdot
denotes $d/dt$.

\narrowtext
It should be noted that the above expression for ${\bf a}_{SO}$ is not
unique; it depends on a ``spin supplementary condition" (SSC) which is related
to the definition of the center-of-mass world line $x_A^\mu$ for each body $A$.
The above form of ${\bf a}_{SO}$ is for the covariant SSC given by
$S_A^{\mu\nu} {u_A}_{\nu} = 0$, where
$u_A^\mu$ is the four-velocity of the center-of-mass world line of body $A$,
and
\end{mathletters}
\begin{equation}
S_A^{\mu\nu} \equiv 2 \int_A (x^{[\mu} - {x_A}^{[\mu}) \tau^{\nu ] 0}
 d^3x , \label{spintensor}
\end{equation}
where $\tau^{\mu\nu}$ denotes the stress-energy tensor of matter plus
gravitational fields satisfying ${\tau^{\mu\nu}}_{,\nu}=0$, and
square brackets around indices denote antisymmetrization.
Note that the spin vector $\bf S$
of each body is defined by
$S_A^i = {1 \over 2} \epsilon_{ijk}S_A^{jk}$.
We discuss the issue of SSCs in more detail in
Appendix \ref{sec:appA}.  Here we simply
wish to emphasize that since we have chosen a center-of-mass world line
for each body through our choice of a SSC, we must ensure that all our
calculations are consistent with this choice.

Since the spin of each body is of order $mR_A\bar v_A$
where $R_A$ is the size of
body $A$ and $\bar v_A$ is its rotational velocity, we see that
the spin-orbit and spin-spin accelerations are of order $(R_A/r)v\bar v_A$ and
$(R_A/r)^2\bar v_A^2$, respectively, compared to the Newtonian acceleration;
these terms thus are formally of (post)$^1$-Newtonian order.  For compact
objects, however, $R_A$ is of order $m$, while $\bar v_A$ could be of order
unity for sufficiently rapid rotation,
so that the spin-orbit and spin-spin accelerations are effectively of
(post)$^{3/2}$-Newtonian and (post)$^2$-Newtonian order, respectively.  If
the bodies are slowly rotating,
the spin contributions to the acceleration
will be even smaller.

It is interesting to note that while ${\bf a}_N$, ${\bf a}_{PN}$,
${\bf a}_{2PN}$, and ${\bf a}_{RR}$ are all confined to the orbital plane,
in general ${\bf a}_{SO}$ and ${\bf a}_{SS}$ are not.  As a result, the
orbital plane will precess in space (except for specific spin orientations)
resulting in modulations of the observed waveform.
We will discuss this effect in more detail in
Sec. \ref{sec:circ}.  Iyer and Will \cite{iyerwill} have
derived post-Newtonian corrections to
${\bf a}_{RR}$ at ${\cal O}(\epsilon^{7/2})$ and
${\cal O}(\epsilon^4)$ where the latter
are the spin-orbit corrections to radiation reaction.

\subsection{Spin Precession Equations}
\label{sec:spinprec}

In addition to the precession of the orbital plane, there are precessions of
the spin vectors themselves.  This effect has been studied by numerous
authors \cite{barker,hartle,damour300}; the relevent equations are
\begin{mathletters}
\begin{eqnarray}
{\bf \dot S_1} = {1 \over r^3} \biggl\{ && ({\bf L_N \times S_1})(2+{3 \over 2}
{m_2 \over m_1}) - {\bf S_2 \times S_1} \nonumber \\ && \mbox{}
+ 3({\bf \hat n \cdot S_2})
{\bf \hat n \times S_1} \biggr\} ,
\end{eqnarray}
\begin{eqnarray}
{\bf \dot S_2} = {1 \over r^3} \biggl\{ && ({\bf L_N \times S_2})(2+{3 \over 2}
{m_1 \over m_2}) - {\bf S_1 \times S_2} \nonumber \\ && \mbox{}
+ 3({\bf \hat n \cdot S_1})
{\bf \hat n \times S_2} \biggr\} ,
\end{eqnarray}
where ${\bf L_N} \equiv \mu ({\bf x \times v})$ is the Newtonian orbital
angular momentum, and
where the first term in each expression is the precession due to spin-orbit
coupling, while the second and third terms
are due to spin-spin coupling.  It is
straightforward to show that the total spin ${\bf S}$ evolves as
\end{mathletters}
\begin{eqnarray}
{\bf \dot S} = {1 \over r^3} \biggl\{ &&
\left[ {\bf L_N \times} \left( {7 \over 2}
{\bf S} + {3 \over 2}{\delta m \over m}{\bf \Delta} \right)
\right] +  3 ({\bf \hat n \cdot S_1})({\bf \hat n
\times S_2}) \nonumber \\ && \mbox{} +
3({\bf \hat n \cdot S_2})({\bf \hat n \times S_1}) \biggr\} .
\label{sdot}
\end{eqnarray}
It is useful to note that the precession of spins is a post-Newtonian effect,
since $L_N/r^3 \approx (v/r)(\mu/r) \approx \epsilon(d/dt)$, and
$S_i/r^3 \approx m_iR\bar v /r^3 \approx \epsilon^{3/2}(d/dt)$.

Since the precession equations have the form ${\bf \dot S_A} =
{\bf \Omega_A \times S_A}$, the
magnitudes of the spins remain constant.  The spin ${\bf S_A}$
instantaneously precesses about the vector ${\bf \Omega_A}$ with a precession
frequency given by $\omega^{(A)}_p = |{\bf \Omega_A}|$.  It is an
instantaneous precession since ${\bf \Omega_A}$ is precessing itself in some
complicated manner.  Notice that if both bodies are spinning, the total spin
${\bf S}$ (with rare exceptions) does not have constant magnitude as the spins
precess.

\subsection{Constants of the Motion}

Through (post)$^2$-Newtonian order, the equations of motion can be
derived from a generalized Lagrangian, that is a Lagrangian which is a
function not just of the relative position and relative velocity, but also
of the relative acceleration \cite{lagrange}.
In our previous papers \cite{isco,kwwspin} we
transformed the Lagrangian into relative coordinates and used it to
compute the energy and total angular
momentum of the system which are conserved to (post)$^2$-Newtonian order,
in the absence of radiation reaction.
Combining our expressions for the non-spinning case and the spinning case, the
energy is given by
\begin{equation}
E =  E_N +  E_{PN} + E_{SO} + E_{2PN} + E_{SS} ,
\end{equation}
where
\widetext
\begin{mathletters}
\begin{equation}
E_N = \mu \left\{ {1 \over 2} v^2 - {m \over r} \right\} ,
\end{equation}
\begin{equation}
E_{PN} =  \mu \left\{
{3 \over 8}
(1-3\eta) v^4
+{1 \over 2} (3+\eta) v^2 {m \over r}  +
{1 \over 2} \eta {m \over r} \dot r^2 + {1 \over 2} ({m \over r})^2 \right\} ,
\end{equation}
\begin{equation}
E_{SO} = {1 \over r^3} {\bf L_N \cdot} ({\bf S} + {\delta m \over m}
{\bf \Delta}) ,
\end{equation}
\begin{eqnarray}
E_{2PN} = \mu \biggl\{ && {5 \over 16}(1-7\eta+13\eta^2) v^6
- {3 \over 8} \eta (1-3\eta){m \over r} \dot r^4
+ {1 \over 8} (21-23\eta-27\eta^2) {m \over r} v^4 \nonumber \\ && \mbox{}
+ {1 \over 8} (14-55\eta+4\eta^2) \left( {m \over r} \right)^2 v^2
+ {1 \over 4} \eta (1-15\eta) {m \over r}
v^2 \dot r^2
 - {1 \over 4} (2+15\eta) \left( {m \over r} \right) ^3 \nonumber \\ && \mbox{}
+ {1 \over 8}
(4+69\eta+12\eta^2) \left( {m \over r} \right) ^2 \dot r^2 \biggr\} ,
\end{eqnarray}
\begin{equation}
E_{SS} = {1 \over r^3} \left\{ 3 \left( {\bf \hat n \cdot S_1} \right)
  \left( {\bf \hat n \cdot S_2} \right) - \left( {\bf S_1 \cdot S_2}
  \right) \right\} ,
\end{equation}
and the total angular momentum is given by
\end{mathletters}
\narrowtext
\begin{equation}
{\bf J} = {\bf L} + {\bf S} ,
\end{equation}
where
\begin{mathletters}
\begin{equation}
{\bf L} = {\bf L}_N + {\bf L}_{PN} + {\bf L}_{SO} + {\bf L}_{2PN} ,
\end{equation}
\begin{equation}
{\bf L}_{PN} = {\bf L_N} \left\{
{1 \over 2} v^2
(1-3\eta) + (3+\eta) {m \over r} \right\} ,
\end{equation}
\begin{eqnarray}
{\bf L}_{SO} = {\mu \over m} \Biggl\{ &&
{m \over r} {\bf \hat n \times} \left[ {\bf \hat n \times} \left(
3 {\bf S} + {\delta m \over m}{\bf \Delta} \right) \right] \nonumber \\
&& \mbox{} - {1 \over 2}
{\bf v \times} \left[ {\bf v \times} \left( {\bf S} + {\delta m \over m}
{\bf \Delta} \right) \right] \Biggr\} ,
\end{eqnarray}
\begin{eqnarray}
{\bf L}_{2PN} = {\bf L_N} \biggl\{ &&
{3 \over 8} (1-7\eta+13\eta^2) v^4
- {1 \over 2}\eta (2+5\eta) {m \over r} \dot r^2 \nonumber \\ && \mbox{}
+ {1 \over 2}
(7-10\eta-9\eta^2) {m \over r}v^2 \nonumber \\ && \mbox{}
+ {1 \over 4} (14-41\eta+4\eta^2)
\left( {m \over r} \right)^2 \biggr\} .
\end{eqnarray}
Note that there is no spin-spin contribution to {\bf J}.
It is straightforward to show that to (post)$^2$-Newtonian order,
$\dot E = {\bf \dot J} = 0,$  where it is understood that whenever the
relative acceleration is found in the time derivative, the
equation of motion carried to the appropriate order is substituted.
\end{mathletters}

\subsection{Precession of the Orbital Angular Momentum}

Since the total angular momentum ${\bf J}$ is conserved (in the absence of
gravitational radiation), it is clear that the orbital angular momentum
${\bf L}$ must precess as
\begin{equation}
{\bf \dot L} = - {\bf \dot S} ,
\end{equation}
where ${\bf \dot S}$ is given by Eq. (\ref{sdot}).

If we restrict ourselves to the case of one spinning body then
\begin{equation}
{\bf \dot S} = {1 \over r^3} \left\{ {1 \over 2} (1 + 3{m \over m_s})
({\bf L_N \times S}) \right\} ,
\end{equation}
where $m_s$ is the mass of the spinning body.
Since to lowest order ${\bf J} = {\bf L_N} + {\bf S}$, then
\begin{equation}
{\bf \dot S} = {1 \over r^3} \left\{ {1 \over 2} (1 + 3{m \over m_s})
({\bf J \times S}) \right\} . \label{onesdot}
\end{equation}
Similarly, since ${\bf L} = {\bf L_N}$ to lowest order,
\begin{equation}
{\bf \dot L} = {1 \over r^3} \left\{ {1 \over 2} (1 + 3{m \over m_s})
({\bf J \times L}) \right\} .
\end{equation}

These two equations imply that ${\bf L}$ and ${\bf S}$ precess about the
fixed vector ${\bf J}$ at the same rate with a precession frequency given by
\begin{equation}
\omega_p = {|{\bf J}| \over 2r^3} \left( 1 + 3{m \over m_s} \right) .
\label{spinprec}
\end{equation}
Note that ${\bf L_N}$ is not necessarily parallel to ${\bf L}$
because of the ${\bf L_{SO}}$ terms, so that the
orbital plane (determined by ${\bf L_N}$)
does not precess in the simple manner above.  Instead the varying
${\bf L_{SO}}$ terms cause it to
wobble slightly on an orbital timescale as it precesses about ${\bf J}$.
This is illustrated in Fig. \ref{wobble}.

In the case of two spinning bodies, ${\bf L}$, ${\bf S_1}$ and ${\bf S_2}$
precess in a very complicated manner (with few exceptions), which can only
be examined numerically.  Fig. \ref{comp_prec_norad}
shows an example of such a precession.
Note that in this case the orbital plane tilts back and forth as it precesses
about ${\bf J}$.
In Sec. \ref{sec:circ} we will examine the precession of the
spins and orbital angular momentum in the case of nearly circular orbits,
and examine the effect of gravitational radiation on the simple precession
for one spinning body, and on the more complicated case of two spinning bodies.

\section{Gravitational Radiation Equations}
\label{sec:rad}

\subsection{Symmetric Trace-Free Radiative Multipoles}

Our goal is to calculate the effects of the spins of the bodies on
the gravitational radiation waveform emitted by the inspiralling binary and on
the energy, angular momentum, and
linear momentum radiated from the system.  Thorne \cite{rmp} showed that
these quantities can be calculated using symmetric trace-free radiative
multipole moments.  For example the gravitational
waveform to (post)$^{3/2}$-Newtonian
order is
given by
\begin{eqnarray}
h^{ij} = {2 \over D} \Biggl\{ && \stackrel{(2)}{I^{ij}} + {1 \over 3}
\stackrel{(3)\,}{I^{ijk}} N^k + {1 \over 12} \stackrel{(4)\quad}
{I^{ijkl}}N^kN^l \nonumber \\ && \mbox{}
+ {1 \over 60} \stackrel{(5)\qquad}{I^{ijklm}}N^kN^lN^m + \cdots
\nonumber \\ && \mbox{}
+ \epsilon^{kl(i} \Biggl[ {4 \over 3} \stackrel{(2)\,}{J^{j)k}}N^l
+ {1 \over 2} \stackrel{(3)\quad}{J^{j)km}}N^lN^m \nonumber \\ && \mbox{}
+ {2 \over 15}
\stackrel{(4)\qquad}{J^{j)kmn}} N^lN^mN^n + \cdots \Biggr] \Biggr\}_{TT} ,
\label{multipolewaveform}
\end{eqnarray}
where $I^{ij\cdots}$ are the mass multipole moments (see below),
$J^{ij\cdots}$ are the current multipole moments, $D$ is the distance from
the source to the observer, $N^i$ is a unit vector from the center of mass
of the source to the observation point, the notation $(n)$ over
each multipole moment denotes the number of derivatives with respect to
retarded
time, $\epsilon^{ijk}$ is the completely antisymmetric Levi-Civita symbol, and
parentheses around indices denote symmetrization. To the accuracy we need,
the mass quadrupole moment $I^{ij}$ needs to be calculated to
(post)$^{3/2}$-Newtonian order beyond the lowest order,
the mass octopole moment $I^{ijk}$ and current
quadrupole moment $J^{jk}$ need to be
calculated to (post)$^1$-Newtonian order, $I^{ijkl}$
and $J^{jkm}$ need to be calculated to (post)$^{1/2}$-Newtonian order, and
$I^{ijklm}$ and $J^{jkmn}$ need to be known only to lowest order.

Blanchet, Damour, and Iyer \cite{bd,di} (BDI)
have developed a formalism for calculating
these radiative multipole moments
in terms of integrals over the source stress-energy.
We use
the BDI formalism to evaluate the spin-orbit and spin-spin corrections to the
radiative multipole moments to the necessary order.

\subsubsection{Mass multipole moments}

The mass multipole moments in harmonic coordinates (the coordinates in which
our equations of motion are written) are given to post-Newtonian order
by Eq. (3.34) in Blanchet and Damour
\cite{bd} as
\begin{eqnarray}
I^L(u) &=& \int (x^L)^{STF} \sigma({\bf x},u)d^3x \nonumber \\ && \mbox{}
- {4(2l+1) \over (l+1)(2l+3)}
{d \over du}\int (x^{iL})^{STF}
\sigma^i({\bf x},u)d^3x
\nonumber \\ && \mbox{}
+ {1 \over 2(2l+3)}
{d^2 \over du^2} \int |{\bf x}^2|(x^L)^{STF} \sigma({\bf x},u)d^3x ,
\nonumber \\
\label{massmultipoles}
\end{eqnarray}
where $L$ denotes a multi-index (i.e. $x^L \equiv x^{i_1}x^{i_2}\dots
x^{i_l}$), the superscript STF denotes that only the symmetric trace-free
part is to be taken,
and the source densities are given by
\begin{mathletters}
\label{sourcet}
\begin{equation}
\sigma({\bf x},t) = T^{00} + T^{ii} ,
\end{equation}
\begin{equation}
\sigma^i({\bf x},t) = T^{0i} .
\end{equation}
\end{mathletters}

In previous calculations involving nonspinning bodies we were able to
evaluate the integrals by assuming a point particle limit.  Taking into
account the spins of the bodies, however, precludes this.
Instead we will assume the bodies to be well-separated,
approximately spherically symmetric (in harmonic coordinates), stationary,
rigidly rotating compact objects whose
structure is given by that of a perfect fluid.  We will then neglect
any effects due to the finite size of the bodies with the exception
of each body's spin.  The stress-energy tensor for a perfect fluid
to the order we need is
\begin{mathletters}
\label{tij}
\begin{equation}
T^{00} = \rho^* (1 + \Pi + \case{1}{2}v^2 - U) ,
\end{equation}
\begin{equation}
T^{0i} = \rho^* v^i ,
\end{equation}
\begin{equation}
T^{ij} = \rho^* v^iv^j + p\delta^{ij} ,
\end{equation}
where $\rho^*=\rho(1+{1 \over 2}v^2 + 3U)$ is the so-called
``conserved density'' (it satisfies a continuity equation to post-Newtonian
order) \cite{tegp}, with $\rho$ the local mass
density, $v$ the velocity, and $U$ the Newtonian gravitational potential;
$\Pi$ is the specific internal energy density, and $p$ is the pressure.
\end{mathletters}

Substituting Eqs. (\ref{sourcet}) and (\ref{tij}) into Eq.
(\ref{massmultipoles}) for the case $l=2$, the mass quadrupole moment is given
by
\begin{eqnarray}
I^{ij} =  \sum_A \Biggl\{ && \int_A (x^ix^j)^{STF} \rho^*({\bf x}) \biggl[
1 + {3 \over 2} v^2 - U
+ 3 {p \over \rho^*({\bf x})}
\nonumber \\ && \mbox{}
+ \Pi \biggr] d^3x
+ {1 \over 14} {d^2 \over dt^2} \int_A (x^ix^j)^{STF} {\bf x}^2 \rho^*
({\bf x}) d^3x \nonumber \\ && \mbox{}
- {20 \over 21}{d \over dt} \int_A (x^ix^jx^k)^{STF} \rho^*
({\bf x}) v^k d^3x \Biggr\} . \label{massquadrupole}
\end{eqnarray}

Following previous post-Newtonian calculations \cite{agw92,bs},
we choose the following provisional definition for the center of mass
for each body:
\begin{equation}
 x_A^i = {1 \over m_A} \int\limits_A x^i \rho^* \left( {\bf x} \right)
  \left[ 1 + {1 \over 2} \bar v_A^2 + \Pi - {1 \over 2} \bar U_A \right]
  d^3x  ,
\label{cm}
\end{equation}
where
\begin{equation}
m_A = \int\limits_A \rho^* \left( {\bf x} \right) \left[ 1 + {1 \over 2}
\bar v_A^2 + \Pi - {1 \over 2} \bar U_A \right] d^3x ,
\end{equation}
where $ \bar v_A^i = v^i - v_A^i$, $v_A^i = dx_A^i /dt$,
and $ \bar U_A $ is the Newtonian potential produced by the $A$-th body
itself.  It turns out that this definition of the center-of-mass world line
does not correspond to the center-of-mass world line of our equations of
motion chosen through the use of a SSC, but rather to one given by a
different SSC.  There does exist a transformation between the two world lines,
given by \cite{sscshift}
\begin{equation}
x_A^i \longrightarrow x_A^i + {1 \over 2m_A} \left( {\bf v_A \times S_A}
  \right)^i , \label{ssccmshift}
\end{equation}
This shift in the world line is of post-Newtonian order, so it can be neglected
at lowest order.  We choose to use our provisional definition of the center of
mass to evaluate the integrals, and then use the transformation on the result
so that it is consistent with our equations of motion.  See Appendix
\ref{sec:appA}
and Ref. \cite{sscshift} for more details.

Eq. (\ref{massquadrupole}) has been evaluated by several authors
\cite{agw92,bs}
for the case of nonspinning bodies, i.e. $\bar v_A = 0$.  By substituting
$x^i = x_A^i + \bar x_A^i$ and $v^i = v_A^i+ \bar v_A^i$ into Eq.
(\ref{massquadrupole}), using the center-of-mass definition Eq. (\ref{cm})
and a virial theorem, and neglecting terms containing $\bar x_A^i \bar x_A^j$
which are ${\cal O}(\beta^2)$ relative to $x_A^i x_A^j$ where
$\beta \equiv R/r$, Blanchet and Sch\"afer \cite{bs} have rewritten Eq.
(\ref{massquadrupole}) as
\widetext
\begin{eqnarray}
I^{ij} &=& \sum_A \Biggl\{ m_A(x_A^ix_A^j)^{STF} \left[ 1 + {3 \over 2} v_A^2
- \sum_{B \neq A} {m_B \over r_{AB}} \right]
- {20 \over 21} {d \over dt} \biggl[
3 \int_A \rho^*({\bf x}) \bar v_A^k
(\bar x_A^ix_A^jx_A^k)^{STF} d^3x \nonumber \\ && \mbox{}
+ m_A (x_A^ix_A^jx_A^k)^{STF} v_A^k \biggr]
+ 6v_A^k \int_A \rho^*({\bf x})(x_A^i \bar x_A^j)^{STF}\bar v_A^k d^3x
+ {1 \over 14} {d^2 \over dt^2}
\left[ m_A x_A^2 (x_A^ix_A^j)^{STF} \right]
\nonumber \\ && \mbox{}
+ {1 \over 2} (x_A^ix_A^j)^{STF} {d^2 \over dt^2} \int_A \rho^*
({\bf x}) \bar x_A^2 d^3x
+ 2 \int_A \rho^* ({\bf x}) (x_A^i\bar x_A^j)^{STF}
\left[ \bar v_A^2 - {1 \over 2} \bar U_A + {3p \over \rho^*({\bf x})}
\right] d^3x
\Biggr\} .
\nonumber \\ && \label{bsmq}
\end{eqnarray}
\narrowtext

The third integral in Eq. (\ref{bsmq}) is just the mass quadrupole moment of
body $A$, which is of ${\cal O}(\beta^2)$ so we will neglect it.
The last integral in Eq. (\ref{bsmq}) will also
vanish because of our assumptions of approximate
spherical symmetry and rigid rotation.
This just leaves integrals of the type:
\begin{eqnarray}
\int_A \rho^*({\bf x}) \bar x_A^a \bar v_A^b d^3x & = &
\int_A \rho^*({\bf x}) \left\{ \bar x_A^{[a} \bar v_A^{b]}
+ \bar x_A^{(a} \bar v_A^{b)} \right\} d^3x  \nonumber \\
& = & {1 \over 2} S_A^{ab} + {1 \over 2} {d \over dt} \int_A \rho^*({\bf x})
\bar x_A^a \bar x_A^b d^3x , \nonumber
\end{eqnarray}
where we have used Eq. (\ref{spintensor}) evaluated to lowest order.
The second term on the right-hand
side vanishes by our assumption of stationary spherical symmetry
so we will neglect it.
Changing the spin tensor to
a spin vector we obtain
\begin{equation}
\int_A \rho^*({\bf x}) \bar x_A^a \bar v_A^b d^3x  =  {1 \over 2}
\epsilon^{iab} S_A^i . \label{spinint}
\end{equation}

Substituting Eq. (\ref{spinint}) into Eq. (\ref{bsmq}) and
carefully counting the
STF parts we obtain
\widetext
\begin{eqnarray}
I^{ij} &=& \sum_A \Biggl\{ m_A(x_A^ix_A^j)^{STF} \left[ 1 + {3 \over 2} v_A^2
- \sum_{B \neq A} {m_B \over r_{AB}} \right]
- {20 \over 21} {d \over dt} \left[ m_A (x_A^ix_A^jx_A^k)^{STF} v_A^k
\right]
\nonumber \\ && \mbox{} + {1 \over 14} {d^2 \over dt^2}
\left[ m_A x_A^2 (x_A^ix_A^j)^{STF} \right]
+ 3 \left[ x_A^i ({\bf v_A
\times S_A})^j \right]^{STF}
- {4 \over 3}{d \over dt} \left[ x_A^i({\bf x_A \times S_A})^j
\right]^{STF} \Biggr\} . \nonumber \\ &&
\end{eqnarray}

Using Eq. (\ref{ssccmshift}) so that we have a consistent center-of-mass
definition with our equations of motion, the mass quadrupole moment becomes
\begin{eqnarray}
I^{ij} &=& \sum_A \Biggl\{ m_A(x_A^ix_A^j)^{STF} \left[ 1 + {3 \over 2} v_A^2
- \sum_{B \neq A} {m_B \over r_{AB}} \right]
- {20 \over 21} {d \over dt} \left[ m_A (x_A^ix_A^jx_A^k)^{STF} v_A^k
\right]
\nonumber \\ && \mbox{} + {1 \over 14} {d^2 \over dt^2}
\left[ m_A x_A^2 (x_A^ix_A^j)^{STF} \right]
+ 4 \left[ x_A^i ({\bf v_A
\times S_A})^j \right]^{STF}
- {4 \over 3}{d \over dt} \left[ x_A^i({\bf x_A \times S_A})^j
\right]^{STF} \Biggr\} . \nonumber \\ &&
\end{eqnarray}
\narrowtext

We rewrite the mass quadrupole moment in relative coordinates by using the
transformations
\begin{mathletters}
\begin{equation}
{\bf x_1} = {\bf x}\left[ {m_2 \over m} + {1 \over 2} \eta {\delta m \over m}
(v^2 - {m \over r}) \right] + {\eta \over m} ({\bf v \times \Delta}) ,
\end{equation}
\begin{equation}
{\bf x_2} = {\bf x}\left[ -{m_1 \over m} +
{1 \over 2} \eta {\delta m \over m}
(v^2 - {m \over r}) \right] + {\eta \over m} ({\bf v \times \Delta}) ,
\end{equation}
which can be obtained from a constant of
the motion that can be taken as the center of mass
\cite{cmshift}.
\end{mathletters}

The relative
mass quadrupole moment to (post)$^{3/2}$-Newtonian order is
\begin{mathletters}
\label{rmmm}
\begin{eqnarray}
I^{ij} &=&  \mu \left( x^i x^j \right)^{STF} \left[ 1 +
{29 \over 42}(1-3\eta)v^2
- {1 \over 7}(5-8\eta){m \over r} \right] \nonumber \\ && \mbox{}
- {4 \over 7}(1-3\eta) \mu r \dot r
(x^iv^j)^{STF} \nonumber \\ && \mbox{}
+ {11 \over 21}(1-3\eta) \mu r^2 (v^iv^j)^{STF}
 + {8 \over 3} \eta \left[ x^i
  \left( {\bf v \times \xi} \right)^j \right]^{STF}
\nonumber \\ && \mbox{} - {4 \over 3} \eta
  \left[ v^i \left( {\bf x \times \xi} \right)^j \right]^{STF}
+ I^{ij}_{Tail} ,
\end{eqnarray}
where ${\bf \xi} \equiv {\bf S} + {\delta m \over m}{\bf \Delta}$, and
where $I^{ij}_{Tail}$ is a reminder that tail effects need to be included
at (post)$^{3/2}$-Newtonian order.  See Wiseman \cite{agwtail} and
Blanchet and Damour \cite{bdtail} for more details on gravitational wave tails.

Note that the spin-orbit correction is a (post)$^{3/2}$-Newtonian correction
for compact objects.  Since a spin-orbit contribution to a multipole
requires a term involving $\bar v_A$, and a spin-spin contribution
$\bar v_A \bar v_B$, it is easy to see that Eq. (\ref{massmultipoles})
implies that there are no spin-orbit corrections to the higher mass
multipole moments at lowest order, and no spin-spin contributions
at lowest order or at post-Newtonian order.
Therefore the remaining mass multipole moments to the appropriate order
are as given by Wiseman \cite{agw92}
\begin{eqnarray}
I^{ijk} =&& - \mu {\delta m \over m} \biggl\{
\left( x^i x^j x^k \right)^{STF} \biggl[ 1
+ {1 \over 6}(5 - 19\eta)v^2 \nonumber \\ && \mbox{}
- {1 \over 6}(5 - 13\eta){m \over r} \biggr]
- (1-2\eta)
r \dot r
(x^ix^jv^k)^{STF} \nonumber \\ && \mbox{} + (1-2\eta)
r^2 (x^iv^jv^k)^{STF} \biggr\},
\end{eqnarray}
\begin{equation}
I^{ijkl}  = \mu (1-3\eta) \left( x^i x^j x^k x^l \right)^{STF},
\end{equation}
\begin{equation}
I^{ijklm}  = - \mu {\delta m \over m} (1-2\eta)
\left( x^i x^j x^k x^l x^m \right)^{STF}.
\end{equation}
\end{mathletters}

\subsubsection{Current multipole moments}

The current multipole moments are given to lowest order as
\begin{equation}
J^{iL} = \left\{ \epsilon^{iab} \int \sigma^b x^{aL} d^3x \right\}^{STF} .
\label{currentmultipoles}
\end{equation}
At leading order, the current quadrupole moment $J^{ij}$ is a moment of angular
momentum density; thus it will give an orbital angular momentum contribution
as well as a spin contribution (effectively at ${\cal O}(\epsilon^{1/2})$).
Although there is also a post-Newtonian correction
to the current quadrupole moment,
spin-orbit terms arising from this correction will be effectively a
(post)$^{3/2}$-Newtonian correction, which is higher than we need, since the
contribution of $J^{ij}$ to the waveform is already
${\cal O}(\epsilon^{1/2})$ at
leading order.

Substituting Eqs. (\ref{sourcet}) and (\ref{tij}) into
Eq. (\ref{currentmultipoles}) we obtain
\begin{equation}
J^{ij} = \left[ \sum_A \epsilon^{iab} \int_A \rho^*{(\bf x)} v^bx^ax^j d^3x
\right]^{STF} , \label{currentquad}
\end{equation}
for the current quadrupole moment.

Substituting $x^i = x_A^i + \bar x_A^i$ and $v^i = v_A^i+ \bar v_A^i$ into Eq.
(\ref{currentquad}), using the center-of-mass definition Eq. (\ref{cm}), and
neglecting terms of ${\cal O}(\beta^2)$, we obtain
\begin{eqnarray}
J^{ij} =  \Biggl\{ \sum_A \epsilon^{iab} \biggl[ && m_Ax_A^ax_A^jv_A^b +
x_A^a \int_A \rho^*({\bf x}) \bar x_A^j \bar v_A^b d^3x \nonumber \\ && \mbox{}
+ x_A^j \int_A \rho^*({\bf x})
\bar x_A^a \bar v_A^b d^3x \biggr] \Biggr\}^{STF}.
\end{eqnarray}
Using Eq. (\ref{spinint}) the current quadrupole moment is then given by
\begin{equation}
J^{ij} = \sum_A \left\{ m_A \left[ x_A^i ({\bf x_A
  \times v_A}) ^j \right]^{STF} + {3 \over 2} \left( x_A^i  S_A^j
\right)^{STF} \right\} . \label{cquad}
\end{equation}
Note that the spin-orbit correction is effectively
(post)$^{1/2}$-Newtonian order.  Repeating the calculation for the
current octopole moment we obtain
\begin{eqnarray}
J^{ijk} =  \sum_A \biggl\{ && m_A \left[ x_A^i x_A^j ({\bf x_A
  \times v_A})^k \right]^{STF} \nonumber \\ && \mbox{}
+ 2 \left( x_A^i x_A^j S_A^k
\right)^{STF} \biggr\}. \label{coct}
\end{eqnarray}

Transforming Eqs. (\ref{cquad}) and (\ref{coct})
into relative coordinates and adding the
post-Newtonian correction to the current quadrupole moment
derived by Wiseman \cite{agw92}, we obtain
\begin{mathletters}
\label{rcmm}
\begin{eqnarray}
J^{ij} &=& - \mu {\delta m \over m} \biggl\{ \left[ x^i ({\bf x
  \times v}) ^j \right]^{STF} \biggl[ 1 + {1 \over 28}(13-68\eta)v^2
\nonumber \\ && \mbox{} + {1 \over 14}(27+30\eta){m \over r} \biggr]
+ {5 \over 28}(1-2\eta) r\dot r \nonumber \\ &&  \times \left[ v^i
({\bf x \times v})^j \right]^{STF}
\biggr\} - {3 \over 2} \eta ( x^i \Delta^j)^{STF},
\end{eqnarray}
\begin{equation}
J^{ijk} = \mu (1-3\eta) \left[ x^i x^j ({\bf x
  \times v})^k \right]^{STF} +  2 \eta (x^i x^j \xi^k)^{STF}.
\end{equation}
The final moment we need is given by
\begin{equation}
J^{ijkl} = - \mu {\delta m \over m} (1-2\eta)
\left[ x^i x^j x^k ({\bf x
  \times v})^l \right]^{STF} .
\end{equation}
Note that there are no spin-spin contributions to the current multipole
moments at this order.
\end{mathletters}

\subsection{Gravitational Waveform}

Taking time derivatives of the radiative multipole moments
(\ref{rmmm}) and (\ref{rcmm}), substituting the
equations of motion (\ref{pneom}) where
appropriate, and substituting the results into
Eq. (\ref{multipolewaveform}), the gravitational waveform is given by
\begin{eqnarray}
h^{ij} = {2 \mu \over D} && \Bigl[ Q^{ij} + P^{0.5}Q^{ij}
+ PQ^{ij} + PQ_{SO}^{ij} + P^{1.5}Q^{ij} \nonumber \\ && \mbox{}
+ P^{1.5}Q_{SO}^{ij} + P^{1.5}Q_{Tail}^{ij}
+ P^2Q_{SS}^{ij} \Bigr]_{TT} , \label{hij}
\end{eqnarray}
where
\widetext
\begin{mathletters}
\begin{equation}
Q^{ij} = 2 \left[ v^iv^j - {m \over r}n^in^j \right] ,
\end{equation}
\begin{equation}
P^{0.5}Q^{ij} = {\delta m \over m}
\left\{ 3 {m \over
r} \left[ 2n^{(i}v^{j)} - \dot r n^in^j \right]  ({\bf \hat N \cdot \hat n})
+ \left[ {m \over r} n^in^j - 2v^iv^j \right]
({\bf \hat N \cdot v}) \right\} ,
\end{equation}
\begin{eqnarray}
PQ^{ij} =&& {1 \over 3} (1-3\eta) \left\{
4{m \over r} \left[ 3\dot r n^in^j - 8n^{(i}v^{j)} \right]
({\bf \hat N \cdot \hat n})({\bf \hat N \cdot v})  + 2
\left[ 3v^iv^j - {m \over r} n^in^j \right] ({\bf \hat N \cdot v})^2
\right. \nonumber \\ && \mbox{} \left.
+ {m \over r} \left[ (3v^2 - 15\dot r^2 + 7{m \over r})n^in^j + 30\dot r
n^{(i}v^{j)} - 14v^iv^j \right]
({\bf \hat N \cdot \hat n})^2
\right\}
+ {4 \over 3} {m \over r} \dot r (5+3\eta)n^{(i}v^{j)} \nonumber \\ && \mbox{}
+ \left[ (1-3\eta)v^2 - {2 \over 3}(2-3\eta){m \over r} \right] v^iv^j +
{m \over r} \left[ (1-3\eta)\dot r^2 - {1 \over 3} (10+3\eta)v^2 +
{29 \over 3} {m \over r} \right] n^in^j , \nonumber \\ &&
\end{eqnarray}
\begin{equation}
PQ_{SO}^{ij} =  {2 \over r^2} ({\bf \Delta \times
\hat N})^{(i} n^{j)} ,
\end{equation}
\begin{eqnarray}
P^{1.5}Q^{ij} =&& {\delta m \over m}(1-2\eta) \Biggl\{
{1 \over 4}{m \over r} \Biggl[ (45 \dot r^2 - 9v^2
-28{m \over r})n^in^j + 58v^iv^j -108 \dot r n^{(i}v^{j)} \Biggr]
({\bf \hat N \cdot \hat n})^2({\bf \hat N \cdot v})
\nonumber \\ && \mbox{}
+ {1 \over 2} \left[ {m \over r}n^in^j
- 4v^iv^j \right] ({\bf \hat N \cdot v})^3
+ {m \over r} \Biggl[ {5 \over 4}(3v^2-7\dot r^2 +6{m \over r})\dot r n^in^j
- {1 \over 6}(21v^2-105\dot r^2
\nonumber \\ && \mbox{}
+44{m \over r})n^{(i}v^{j)}
- {17 \over 2} \dot r v^iv^j \Biggr] ({\bf \hat N \cdot \hat n})^3
+{3 \over 2} {m \over r} \left[ 10n^{(i}v^{j)} - 3 \dot r n^in^j \right]
({\bf \hat N \cdot \hat n})({\bf \hat N \cdot v})^2
\Biggr\}
\nonumber \\ && \mbox{}
+ {\delta m \over m} {1 \over 12} {m \over r}({\bf \hat N \cdot \hat n})
\Biggl\{  n^in^j\dot r \left[ \dot r^2(15
-90\eta)-v^2(63-54\eta)+{m \over r}(242-24\eta) \right]
\nonumber \\ &&
-\dot r v^iv^j(186+24\eta)
+ 2n^{(i}v^{j)} \left[ \dot r^2 (63+54\eta) - {m \over r}(128-36\eta)
+ v^2(33-18\eta) \right] \Biggr\}
\nonumber \\ && \mbox{}
+ {\delta m \over m} ({\bf \hat N \cdot v}) \Biggl\{
{1 \over 2}v^iv^j \left[ {m \over r}(3-8\eta)-2v^2(1-5\eta) \right]
-n^{(i}v^{j)}{m \over r}\dot r (7+4\eta) \nonumber \\ &&
- n^in^j{m \over r} \left[ {3 \over 4}(1-2\eta)\dot r^2
+ {1 \over 3}(26-3\eta){m \over r} - {1 \over 4}(7-2\eta)v^2 \right]
\Biggr\} ,
\end{eqnarray}
\begin{eqnarray}
P^{1.5}Q_{SO}^{ij} =&&
{2 \over r^2} \Biggl\{ n^in^j \left[ ({\bf \hat n
\times v}){\bf \cdot}(12{\bf S} +6 {\delta m \over m}{\bf \Delta})\right]
- n^{(i}\left[{\bf v \times}(9{\bf S}+5{\delta m \over m}{\bf \Delta})\right]
^{j)}  \nonumber \\ && \mbox{}
+ \left[ 3 \dot r ({\bf \hat N \cdot \hat n})
- 2  ({\bf \hat N \cdot v}) \right] \left[ ({\bf S}+{\delta m \over m}
{\bf \Delta}){\bf \times \hat N} \right]^{(i} n^{j)}
- v^{(i}\left[{\bf \hat n \times}(2{\bf S}+2{\delta m \over m}{\bf \Delta})
\right]^{j)} \nonumber \\ && \mbox{}
+ \dot r n^{(i}\left[{\bf \hat n \times}
(12{\bf S}+6{\delta m \over m}{\bf \Delta})
\right]^{j)}
- 2  ({\bf \hat N \cdot \hat n}) \left[ ({\bf S}+{\delta m \over m}
{\bf \Delta}){\bf \times \hat N} \right]^{(i} v^{j)} \Biggr\} ,
\end{eqnarray}
\begin{equation}
P^{1.5}Q_{Tail}^{ij}  = 2 {m \over \mu} \int_0^\infty \stackrel{(4)}{I_N^{ij}}
(u-u') \left[ \ln \left( {u' \over 2s} \right) + {11 \over 12} \right] du' ,
\end{equation}
\begin{equation}
P^2Q_{SS}^{ij} = - {6 \over \mu r^3} \left\{ n^i n^j
\left[ ({\bf S_1 \cdot S_2})
- 5 ({\bf \hat n \cdot S_1})({\bf \hat n \cdot S_2}) \right] +
2n^{(i}S_1^{j)} ({\bf \hat n \cdot S_2}) +
2n^{(i}S_2^{j)} ({\bf \hat n \cdot S_1}) \right\} ,
\end{equation}
where $Q^{ij}$ is just the standard quadrupole term, $P^{0.5}Q^{ij}$ and
$PQ^{ij}$ were derived by Wagoner and Will \cite{wagwill}, $P^{1.5}Q^{ij}$ was
derived by Wiseman \cite{agw92}, and $PQ_{SO}^{ij}$ and $P^{1.5}Q_{SO}^{ij}$
are the explicit spin-orbit corrections to the waveform.  $P^{1.5}Q_{Tail}^
{ij}$ is the leading order contribution to the waveform due to the tail,
where $I_N^{ij} = \mu \left( x^ix^j \right)^{STF}$.
Notice that it depends upon the past history of the binary's inspiral
($u$ is a retarded time), and that $s$ is an arbitrary matching parameter.
See Wiseman \cite{agwtail} and Blanchet and Damour \cite{bdtail} for more
details about gravitational wave tails.
Note that we have included the leading order spin-spin contribution to the
waveform even though it is effectively a (post)$^2$-Newtonian
term.  It is due entirely
to substituting ${\bf a_{SS}}$ into the time derivatives of the mass
quadrupole moment.  Note that we have simplified Eq. (\ref{hij})
by using relations
such as
\end{mathletters}
\narrowtext
\[
\left\{ \epsilon^{kl(i} \left[ a^{j)} b^k \right]^{STF} N^l \right\}_{TT}
= \left[ \left( {\bf b \times \hat N} \right)^{(i} a^{j)} \right]_{TT}.
\]
Fig. \ref{gravitational_waveform}
shows the different contributions to the waveform for the final
few orbits of an inspiralling binary system.  We see that the spin-orbit
contribution is comparable to the other post-Newtonian contributions, but
the spin-spin contribution is almost negligible.

\subsection{Energy Loss}

The radiative energy loss in terms of STF radiative multipoles is given by
Thorne \cite{rmp} as
\begin{equation}
{dE \over dt} = - {1 \over 5} \left\{ \stackrel{(3)}{I_{ij}}
 \stackrel{(3)}{I_{ij}} + {5 \over 189} \stackrel{(4)}{I_{ijk}}
 \stackrel{(4)}{I_{ijk}} + {16 \over 9} \stackrel{(3)}{J_{ij}}
\stackrel{(3)}{J_{ij}} \right\},
\end{equation}
for the accuracy we require.
Taking time derivatives of the radiative multipole moments
(\ref{rmmm}) and (\ref{rcmm}), and substituting the
equations of motion (\ref{pneom}) where
appropriate, the energy loss
is given by
\begin{equation}
{dE \over dt} = {\dot E}_N + {\dot E}_{PN} + {\dot E}_{SO} + {\dot E}_{Tail}
+ {\dot E}_{SS},
\end{equation}
where
\begin{mathletters}
\begin{equation}
{\dot E}_N  = - {8 \over 15} {m^2 \mu^2 \over r^4} \left\{ 12v^2 - 11 \dot
r^2 \right\},
\end{equation}
\begin{eqnarray}
{\dot E}_{PN}  =&& - {2 \over 105} {m^2 \mu^2 \over r^4} \biggl\{
(785-852\eta)v^4 -160(17-\eta) {m \over r}v^2 \nonumber \\ && \mbox{}
+ 8(367-15\eta){m \over r} \dot r^2
-2(1487-1392\eta)v^2 \dot r^2 \nonumber \\ && \mbox{}
+ 3(687-620\eta) \dot r^4
+ 16(1-4\eta)({m \over r})^2 \biggr\},
\end{eqnarray}
\begin{eqnarray}
{\dot E}_{SO}  = && - {8 \over 15} {m \mu \over r^6} \Bigg\{ {\bf L_N \cdot}
\Bigg[ {\bf S} (78 \dot r^2 -80v^2 -8
{m \over r}) \nonumber \\ && \mbox{}
+ {\delta m \over m}{\bf \Delta} (51 \dot r^2 - 43v^2 +
4{m \over r} \Biggr] \Biggr\},
\end{eqnarray}
\begin{equation}
{\dot E}_{Tail}  = -{2 \over 5} \mu \stackrel{(3)}{I_N^{ij}} {d \over du}
P^{1.5}Q_{Tail}^{ij} ,
\end{equation}
\begin{eqnarray}
{\dot E}_{SS}  =&& - {4 \over 15} {m \mu \over r^6} \biggl\{
- 3 ({\bf \hat n \cdot S_1})({\bf \hat n \cdot S_2} )
\left( 168 v^2 - 269 \dot r ^2 \right) \nonumber \\ && \mbox{} +
3 ({\bf S_1 \cdot S_2})  \left( 47 v^2- 55 \dot r ^2 \right)
+ 71 ( {\bf v \cdot S_1})({\bf v \cdot S_2} ) \nonumber \\ && \mbox{}
- 171 \dot r \left[ ({\bf v \cdot S_1})({\bf \hat n \cdot
S_2}) + ({\bf \hat n \cdot S_1})(
  {\bf v \cdot S_2}) \right] \biggr\} , \nonumber \\ &&
\end{eqnarray}
where ${\dot E}_{PN}$ was found by Wagoner and Will \cite{wagwill}
, and ${\dot E}_
{SO}$ and ${\dot E}_{SS}$ were reported in our previous paper \cite{kwwspin}.
$\dot E_{Tail}$ depends upon the past history of the system and can only
be evaluated explicitly for simple cases.  Notice that
the spin-spin contribution to the energy loss, which is effectively a
(post)$^2$-Newtonian correction, comes from using post-Newtonian
equations of motion in the derivatives of the mass quadrupole, and also
from the contraction of the current quadrupoles.  We have ignored
(spin)$^2$-terms which are the same order as the spin-spin terms.
In Fig. \ref{luminosity}
we compare the spin contributions to the energy loss with the
other contributions for an inspiralling binary system.
Again we see that the spin-orbit contribution can be significant, while
the spin-spin contribution is almost negligible.
\end{mathletters}

\subsection{Angular Momentum Loss}

The radiative angular momentum loss in terms of STF radiative multipoles
is given by Thorne \cite{rmp} as
\begin{equation}
{dJ^i \over dt} = - {2 \over 5} \epsilon_{ijk} \left\{ \stackrel{(2)}{I_{jl}}
 \stackrel{(3)}{I_{kl}} + {5 \over 126} \stackrel{(3)}{I_{jlm}}
 \stackrel{(4)}{I_{klm}} + {16 \over 9} \stackrel{(2)}{J_{jl}}
\stackrel{(3)}{J_{kl}} \right\},
\end{equation}
for the accuracy we require.
Taking time derivatives of the radiative multipole moments
(\ref{rmmm}) and (\ref{rcmm}), and substituting the
equations of motion (\ref{pneom}) where
appropriate, the
angular momentum loss is given by
\begin{equation}
{d{\bf J} \over dt} = {\dot {\bf J}}_N + {\dot {\bf J}}_{PN}
+ {\dot {\bf J}}_{SO} + {\dot {\bf J}}_{Tail} + {\dot {\bf J}}_{SS},
\end{equation}
where
\widetext
\begin{mathletters}
\begin{equation}
{\dot {\bf J}}_N  = - {8 \over 5} {m \mu \over r^5} {\bf L_N}
\left\{ 2v^2 - 3 \dot r^2 + 2 {m \over r} \right\},
\end{equation}
\begin{eqnarray}
{\dot {\bf J}}_{PN}  = - {2 \over 105} {m \mu \over r^5}  {\bf L_N}
\biggl\{ && (307-548\eta)v^4 - 6(74-277\eta)v^2 \dot r^2
+ 2(372+197\eta){m \over r} \dot r^2
\nonumber \\ && \mbox{}
+ 15(19-72\eta) \dot r^4
- 4(58+95\eta) {m \over r}v^2
- 2(745-2\eta)({m \over r})^2 \biggr\},
\end{eqnarray}
\begin{eqnarray}
{\bf \dot J}_{SO}  =&& -{4 \over 5} {\mu^2 \over r^3} \left\{
{2 \over 3}{m \over r}(\dot r^2 - v^2){\delta m \over m}{\bf \Delta}
- \dot r {m \over r}{\bf \hat n \times} \left[ 4({\bf v \times S}) +
{5 \over 3}{\delta m \over m}({\bf v \times \Delta}) \right]
\right. \nonumber \\ &&
+ {m \over r}{\bf \hat n \times} \left[ ({\bf \hat n \times S})(15 \dot r^2
-{41 \over 3}v^2 +{4 \over 3}{m \over r}) + {\delta m \over m}
({\bf \hat n \times \Delta}) (9\dot r^2 - 8v^2 -{2 \over 3}{m \over r}) \right]
\nonumber \\ &&
+ \dot r {\bf v \times} \left[ ({\bf \hat n \times S})(18 {m \over r}
+ 44v^2 -55 \dot r^2) + 5 {\delta m \over m}
({\bf \hat n \times \Delta}) ({5 \over 3}{m \over r} +4v^2-5\dot r^2) \right]
\nonumber \\ &&
+ {\bf v \times} \left[ ({\bf v \times S})(36 \dot r^2
-{71 \over 3}v^2 -{50 \over 3}{m \over r}) + {\delta m \over m}
({\bf v \times \Delta}) (18\dot r^2 - {35 \over 3}v^2-9{m \over r}) \right]
\nonumber \\ && \left. + {{\bf L_N}
\over \mu^2 r^2} {\bf L_N \cdot} \left[ (65 \dot r^2 -37v^2
-{163 \over 3}{m \over r}){\bf S} + (35 \dot r^2 -19v^2 -{71 \over 3}
{m \over r}){\delta m \over m}{\bf \Delta} \right] \right\} ,
\end{eqnarray}
\begin{equation}
{\dot {\bf J}}_{Tail}  = -{2 \over 5} \mu \epsilon^{ijk} \left\{
\stackrel{(3)}{I_N^{kl}} P^{1.5}Q_{Tail}^{jl} + \stackrel{(2)}{I_N^{jl}}
{d \over du} P^{1.5}Q_{Tail}^{kl} \right\},
\end{equation}
\begin{equation}
{\dot {\bf J}}_{SS} =  -{2 \over 5} \mu \epsilon^{ijk} \left\{
\stackrel{(2)}{I_N^{jl}}({\bf a_N}) \stackrel{(3)}{I_N^{kl}}({\bf a_{SS}}) +
\stackrel{(2)}{I_N^{jl}}({\bf a_{SS}}) \stackrel{(3)}{I_N^{kl}}({\bf a_N}) +
{16 \over 9}\stackrel{(2)}{J_{SO}^{jl}}({\bf a_N})
\stackrel{(3)}{J_{SO}^{kl}}({\bf a_N}) \right\},
\end{equation}
where ${\dot {\bf J}}_{PN}$ was calculated by Junker and Sch\"afer
\cite{junker}.  Note that ${\dot {\bf J}}_{Tail}$ depends upon the
past history of the system.  To avoid lengthy expressions, we
have left ${\dot {\bf J}}_{SS}$ in terms
of derivatives of the multipole moments, where we have specified which
contribution to the equations of motion should be substituted for the
accelerations which appear in the time derivatives.
Note that while ${\dot {\bf J}}_N$ and ${\dot {\bf J}}_{PN}$ are in the
direction of ${\bf L_N}$, in general ${\dot {\bf J}}_{SO}$ and ${\dot {\bf
J}}_{SS}$ are not.
\end{mathletters}
\narrowtext

\subsection{Linear Momentum Loss}

The radiative linear momentum loss in terms of STF radiative multipoles
is given by Thorne \cite{rmp} as
\begin{equation}
{dP^i \over dt} = - \left\{ {2 \over 63} \stackrel{(4)}{I_{ijk}}
 \stackrel{(3)}{I_{jk}} + {16 \over 45} \epsilon_{ijk} \stackrel{(3)}{I_{jl}}
\stackrel{(3)}{J_{kl}} \right\},
\end{equation}
for the accuracy we require.
Taking time derivatives of the radiative multipole moments
(\ref{rmmm}) and (\ref{rcmm}), and substituting the
equations of motion (\ref{pneom}) where
appropriate, the
linear momentum loss is given by
\begin{equation}
{d{\bf P} \over dt} = {\dot {\bf P}}_N + {\dot {\bf P}}_{SO} ,
\end{equation}
where
\begin{mathletters}
\begin{eqnarray}
{\dot {\bf P}}_N = && - {8 \over 105} {\delta m \over m}
\eta^2 \left( {m \over r}
\right)^4 \biggl\{ \dot r {\bf \hat n} \left[ 55v^2 -45\dot r^2 + 12 {m \over
r}
\right] \nonumber \\ && \mbox{}
+ {\bf v} \left[ 38\dot r^2 - 50v^2 - 8 {m \over r} \right] \biggr\},
\end{eqnarray}
\begin{eqnarray}
{\dot {\bf P}}_{SO} = && - {8 \over 15} {\mu^2 m \over r^5}
\Bigl\{ 4 \dot r ({\bf v \times \Delta})
- 2v^2 ({\bf \hat n \times \Delta}) \nonumber \\ && \mbox{}
- ({\bf \hat n \times v}) \left[ 3\dot r ({\bf \hat n \cdot \Delta})
+ 2 ({\bf v \cdot \Delta}) \right] \Bigr\},
\end{eqnarray}
where ${\dot {\bf P}}_N$ was studied by Fitchett \cite{fitchett}.  Note that
${\dot {\bf P}}_{SO}$, which is effectively a (post)$^{1/2}$-Newtonian
correction, can be directed out of the orbital plane.  There is
no spin-spin contribution to the linear momentum loss at this order.
Wiseman \cite{agw92} has derived the
post-Newtonian corrections to $\dot {\bf P}$.
In Fig. \ref{momentum_ejection} we plot the
momentum ejected for a typical inspiral.  Notice that even in the
presence of spinning bodies, the momentum ejection is periodic, and
therefore there is no large buildup of momentum ejected in a specific
direction.
\end{mathletters}

\section{Circular Orbits}
\label{sec:circ}

\subsection{Circular Orbit Limit}

Gravitational radiation tends to circularize the orbit of an inspiralling
binary.  Therefore we would like to examine the
last several minutes of the inspiral
with the assumption that the orbit is quasi-circular, that is, the orbit
is circular on an orbital timescale, but inspirals on a radiation-reaction
timescale.  This is a
reasonable assumption, since Lincoln and Will \cite{linc}
have shown that virtually
all captured binaries will have sufficient time to circularize their orbits
before plunging to coalescence.

The equations of motion can be rewritten using the identities
\begin{mathletters}
\label{eomident}
\begin{equation}
{\bf \hat n \cdot a}  = \ddot r - r\omega^2 ,
\end{equation}
\begin{equation}
{\bf \hat \lambda \cdot a}  = r \dot \omega + 2 \dot r \omega ,
\end{equation}
\begin{equation}
{\bf \hat L_N \cdot a}  = - r \omega \left( {\bf \hat \lambda \cdot}
{d {\bf \hat L_N} \over dt} \right) ,
\end{equation}
where ${\bf \hat \lambda} = {\bf \hat L_N \times \hat n}$, ${\bf \hat L_N} =
{\bf L_N}/|{\bf L_N}|$, and the orbital angular velocity
$\omega$ is defined by ${\bf v} = \dot r
{\bf \hat n} + r\omega {\bf \hat \lambda}$.  A circular orbit
on a fixed plane is given
by the solution $\ddot r = \dot r = \dot \omega = (d{\bf \hat L_N}/dt)
= 0$.
This solution exists if $r\omega^2 =
- {\bf \hat n \cdot a}$, ${\bf \hat \lambda \cdot a} = 0$, and
${\bf \hat L_N \cdot a} = 0$, where we have substituted $\dot r = 0$ into
the righthand sides of Eqs. (\ref{eomident}).   Examining the equations of
motion
(\ref{pneom}) (while ignoring the radiation-reaction terms), we see that
circular orbit solutions exist only if the spins are aligned perpendicular
to the orbital plane.  If we instead define a circular orbit as one that
has a constant orbital separation, but allow the orbital plane to precess,
there exist circular orbit solutions for the case of one spinning body, but
not for the case of both bodies spinning with general orientations.
\end{mathletters}

If we first average the spin-orbit and spin-spin terms in
the acceleration over an orbit, then we can obtain
orbits of constant separation for arbitrary spins and orientations.
In order to average over an orbit we
need to assume that the spins and orbital plane remain constant over an orbital
period, in other words, that $\omega_p / \omega$ is small.
For circular orbits
\begin{equation}
m \omega = \left( {m \over r} \right)^{3/2} ,
\end{equation}
to leading order.  Using Eq. (\ref{spinprec}) for the case of one
spinning body we see that
\begin{equation}
{\omega_p \over \omega} = {1 \over 2} {|{\bf J}| \over m^2} \left( {m \over r}
\right)^{3/2} \left[ 1 + 3 {m \over m_s} \right].
\end{equation}
Since $|{\bf J}| \leq |{\bf L}| + |{\bf S}|$, $|{\bf L}|/m^2 =
\eta (r/m)^{1/2}$, and $|{\bf S}|/m^2 = \chi (m_s/m)^2$, where $\chi \leq 1$,
then we see that
\begin{eqnarray}
{\omega_p \over \omega} \leq && {1 \over 2} \Biggl\{ \left( {m \over r}
\right) \left[ \eta + 3{m-m_s \over m} \right] \nonumber \\ && \mbox{}
+ \chi \left( {m \over r}
\right)^{3/2} {m_s \over m} \left[ 3 + {m_s \over m} \right] \Biggr\} ,
\end{eqnarray}
which is small ($\approx {\cal O}(m/r)$) until the very
late stages of the inspiral, where the whole
circular orbit approximation breaks down anyway.  We would expect a
similar argument to hold for the case where both bodies are spinning,
since the spins' instantaneous precessions have a form similar to that
of the precession in the single spin case (see Sec. \ref{sec:spinprec}).
An examination of
numerical evolutions of the precession equations confirms this.

In examining circular orbits, we will assume an orbit where $\dot r = 0$
and $r\omega^2 = - \langle {\bf \hat n \cdot a} \rangle$ where the brackets
denote an average over an orbit.
We therefore obtain the following expressions for a circular orbit.  The
orbital velocity is given by $v=r\omega$ where
\widetext
\begin{eqnarray}
r^2 \omega^2  = \left( {m \over r} \right)
\Biggl\{ && 1 - (3-\eta)\left( {m \over r} \right)
- \sum_{i = 1,2} \left[ \chi_i ({\bf \hat L_N \cdot \hat s_i}) ( 2{m_i^2
\over m^2} +3\eta) \right]
\left( {m \over r} \right)^{3/2}
+ \biggl[ (6 + {41 \over 4}\eta + \eta^2) \nonumber \\ && \mbox{}
- {3 \over 2}
\eta \chi_1 \chi_2 \Bigl[ ({\bf \hat s_1 \cdot \hat s_2})
- 3({\bf \hat L_N
\cdot \hat s_1})({\bf \hat L_N \cdot \hat s_2}) \Bigr] \biggr]
\left( {m \over r} \right)^2 \Biggr\},
\end{eqnarray}
where ${\bf S_A} = \chi_A m_A^2 {\bf \hat s_A}$.  The energy and angular
momentum are given by
\begin{eqnarray}
E &=& - {1 \over 2} {\mu m \over r}
\Biggl\{ 1 - {1 \over 4}(7-\eta)\left( {m \over r} \right)
\mbox{} + \sum_{i = 1,2} \left[ \chi_i ({\bf \hat L_N \cdot \hat s_i}) (
2{m_i^2
\over m^2} +\eta) \right]
\left( {m \over r} \right)^{3/2} \nonumber \\ && \mbox{}
- \biggl[ {1 \over 8}(7 - 49\eta - \eta^2)
- {1 \over 2}
\eta \chi_1 \chi_2 \Bigl[ ({\bf \hat s_1 \cdot \hat s_2})
- 3({\bf \hat L_N
\cdot \hat s_1})({\bf \hat L_N \cdot \hat s_2}) \Bigr] \biggr]
\left( {m \over r} \right)^2 \Biggr\}, \label{circe}
\end{eqnarray}
\begin{eqnarray}
{\bf J}  &=& \mu (mr)^{1/2} {\bf \hat L_N} \Biggl\{ 1 + 2\left( {m \over r}
\right) - {1 \over 4}
\sum_{i = 1,2} \left[ \chi_i ({\bf \hat L_N \cdot \hat s_i}) ( 8{m_i^2
\over m^2} +7\eta) \right]
\left( {m \over r} \right)^{3/2}
+ \biggl[ {1 \over 2}(5 -9\eta) \nonumber \\ && \mbox{}
- {3 \over 4}
\eta \chi_1 \chi_2 \left[ ({\bf \hat s_1 \cdot \hat s_2}) - 3({\bf \hat L_N
\cdot \hat s_1})({\bf \hat L_N \cdot \hat s_2}) \right] \biggr]
\left( {m \over r} \right)^2 \Biggr\} \nonumber \\ && \mbox{}
+ {\bf S} - {1 \over 4} \mu (mr)^{1/2}
\sum_{i = 1,2} \left[ \chi_i {\bf \hat s_i} ( 4{m_i^2
\over m^2} +\eta) \right]
\left( {m \over r} \right)^{3/2} .
\end{eqnarray}

For a circular orbit, the waveform is given by
\begin{equation}
h^{ij} = {2 \mu \over D} \left( {m \over r} \right)  \left\{
Q_c^{ij} + P^{0.5}Q_c^{ij} \left( {m \over r} \right)^{1/2}
+ PQ_c^{ij} \left( {m \over r} \right)
+ P^{1.5}Q_c^{ij} \left( {m \over r} \right)^{3/2} \right\}_{TT} ,
\label{circh}
\end{equation}
where
\begin{mathletters}
\begin{equation}
Q_c^{ij} = 2 \left[ \lambda^i\lambda^j - n^in^j \right] ,
\end{equation}
\begin{equation}
P^{0.5}Q_c^{ij} = = {\delta m \over m}
\left\{ 6({\bf \hat N \cdot \hat n})
n^{(i}\lambda^{j)}  + ({\bf \hat N \cdot \hat \lambda})
\left[  n^in^j - 2\lambda^i\lambda^j \right] \right\} ,
\end{equation}
\begin{eqnarray}
PQ_c^{ij} =&&
{2 \over 3} (1-3\eta) \biggl\{ ({\bf \hat N \cdot \hat n}
)^2  \left[ 5n^in^j
- 7\lambda^i\lambda^j \right]
- 16 ({\bf \hat N \cdot \hat n})
({\bf \hat N \cdot \hat \lambda})
n^{(i}\lambda^{j)} \nonumber \\ && \mbox{} +
({\bf \hat N \cdot \hat \lambda})^2
\left[ 3\lambda^i\lambda^j - n^in^j \right] \biggr\}
+ {1 \over 3}
(19-3\eta)(n^in^j- \lambda^i\lambda^j)
+  {2 \over m^2}  n^{(i} ({\bf \Delta \times
\hat N})^{j)}  ,
\end{eqnarray}
\begin{eqnarray}
P^{1.5}Q_c^{ij}  =&& {\delta m \over m}
\biggl\{ (1-2\eta) \biggl[ {1 \over 2} ({\bf \hat N \cdot \hat \lambda})^3
\left( n^in^j -
4\lambda^i\lambda^j \right)
+{1 \over 4} ({\bf
\hat N \cdot \hat n})^2({\bf \hat N \cdot \hat \lambda})
\left( 58\lambda^i\lambda^j
-37n^in^j \right)
\nonumber \\ && \mbox{}
- {65 \over 6} ({\bf \hat N
\cdot \hat n})^3
n^{(i}\lambda^{j)}
+ 15({\bf \hat N \cdot \hat n})({\bf \hat N \cdot
\hat \lambda})^2 n^{(i}\lambda^{j)} \biggr]
-  ({\bf \hat N \cdot \hat \lambda})  \biggl[
{1 \over 12}(101-12\eta) n^in^j
\nonumber \\ && \mbox{}
- {1 \over 2} (19 - 4\eta)\lambda^i\lambda^j \biggr]
- {1 \over 6}(149-6\eta)({\bf \hat N \cdot \hat n})n^{(i}\lambda^{j)} \biggr\}
- {2 \over m^2}   \Biggl\{
\lambda^i\lambda^j \left[ {\bf \hat L_N \cdot} (5{\bf S} + 3{\delta m \over
m}{\bf \Delta}) \right]
\nonumber \\ && \mbox{}
-6n^in^j \left[ {\bf \hat L_N \cdot}(2{\bf S}
+{\delta m \over m}{\bf \Delta})\right]
+ 2 \lambda^{(i}\left[{\bf \hat n \times}({\bf S}
+{\delta m \over m}{\bf \Delta})\right]^{j)}
+  n^{(i}\left[{\bf \hat \lambda \times}(9{\bf S}
+5{\delta m \over m}{\bf \Delta})\right]^{j)}
\nonumber \\ && \mbox{}
+ 2  ({\bf \hat N \cdot \hat \lambda}) \left[ ({\bf S}+{\delta m \over m}
{\bf \Delta}){\bf \times \hat N} \right]^{(i} n^{j)}
+ 2  ({\bf \hat N \cdot \hat n}) \left[ ({\bf S}+{\delta m \over m}
{\bf \Delta}){\bf \times \hat N} \right]^{(i} \lambda^{j)} \Biggr\} ,
\end{eqnarray}
where $\lambda^i = v^i/|{\bf v}|$ for a circular orbit.  The energy loss and
angular momentum loss for a circular orbit are given by
\end{mathletters}
\begin{eqnarray}
{dE \over dt} &=& - {32 \over 5} \eta^2 \left( {m \over r} \right)^5
\Biggl\{ 1 - {1 \over 336}(2927+420\eta)\left( {m \over r} \right)
- \Biggl[ {1 \over 12}
\sum_{i = 1,2}  \biggl[ \chi_i ({\bf \hat L_N
\cdot \hat s_i}) ( 73{m_i^2
\over m^2} +75\eta) \biggr] \nonumber \\ && \mbox{} - 4\pi \Biggr]
\left( {m \over r} \right)^{3/2}
- {1 \over 48}
\eta \chi_1 \chi_2 \left[ 223({\bf \hat s_1 \cdot \hat s_2}) - 649({\bf \hat
L_N
\cdot \hat s_1})({\bf \hat L_N \cdot \hat s_2}) \right]
\left( {m \over r} \right)^2 \Biggr\} , \label{circedot}
\end{eqnarray}
\begin{eqnarray}
{d {\bf J} \over dt}  &=& - {32 \over 5} \eta^2 \left( {m \over r} \right)^4
(mr)^{1/2} \Biggl\{ {\bf \hat L_N} \Biggl[ 1 - {1 \over 336}
(2423+588\eta)\left( {m \over r} \right)
- \biggl( {1 \over 8}
\sum_{i = 1,2}  \biggl[ \chi_i ({\bf \hat L_N \cdot \hat s_i}) ( 53{m_i^2
\over m^2}
\nonumber \\ && \mbox{} + 52\eta) \biggr] - 4\pi \biggr)
\left( {m \over r} \right)^{3/2} \Biggr]
+ {1 \over 24}
\sum_{i = 1,2} \left[ \chi_i {\bf \hat s_i} ( 37{m_i^2
\over m^2} +42\eta) \right]
\left( {m \over r} \right)^{3/2} \Biggr\},
\end{eqnarray}
where the $4\pi$ terms are due to the gravitational wave tail.

The rate of inspiral is given by $\dot r = (dE/dt)/(dE/dr)$.  Taking
Eq. (\ref{circedot}) and dividing it by the derivative of Eq. (\ref{circe}),
we obtain
\begin{eqnarray}
{dr \over dt}  &=& - {64 \over 5} \eta \left( {m \over r} \right)^3
\Biggl\{ 1 - {1 \over 336}(1751+588\eta)\left( {m \over r} \right)
- \Biggl[ {7 \over 12} \sum_{i = 1,2} \biggl[ \chi_i ({\bf \hat L_N
\cdot \hat s_i}) ( 19{m_i^2
\over m^2} +15\eta) \Biggr] \nonumber \\ && \mbox{} - 4\pi \Biggr]
\left( {m \over r} \right)^{3/2}
- {5 \over 48}
\eta \chi_1 \chi_2 \left[ 59({\bf \hat s_1 \cdot \hat s_2}) - 173({\bf \hat L_N
\cdot \hat s_1})({\bf \hat L_N \cdot \hat s_2}) \right]
\left( {m \over r} \right)^2 \Biggr\} . \label{inspiral}
\end{eqnarray}

The above expressions can be inverted to express everything in terms of
the orbital frequency.  For a given orbital frequency, the separation is
given by
\begin{eqnarray}
(r/m)  &=& (m\omega)^{-2/3}
\Biggl\{ 1 - {1 \over 3}(3-\eta) (m\omega)^{2/3}
- {1 \over 3} \sum_{i = 1,2} \left[ \chi_i ({\bf \hat L_N
\cdot \hat s_i}) ( 2{m_i^2
\over m^2} +3\eta) \right]
(m\omega) \nonumber \\ &&  + \left[
\eta({19 \over 4} + {1 \over 9}\eta)
- {1 \over 2}
\eta \chi_1 \chi_2 \left[ ({\bf \hat s_1 \cdot \hat s_2}) - 3({\bf \hat L_N
\cdot \hat s_1})({\bf \hat L_N \cdot \hat s_2}) \right] \right]
(m\omega)^{4/3} \Biggr\} . \label{circr}
\end{eqnarray}
Using Eqs. (\ref{inspiral}) and (\ref{circr}), we find that
the evolution of the orbital frequency is given by
\begin{eqnarray}
{{\dot \omega} \over \omega^2}  &=& {96 \over 5} \eta (m\omega)^{5/3}
\Biggl\{ 1 - {1 \over 336}(743+924\eta) (m\omega)^{2/3}
- \Biggl[ {1 \over 12} \sum_{i = 1,2} \biggl[ \chi_i ({\bf \hat L_N
\cdot \hat s_i}) ( 113{m_i^2
\over m^2}  +75\eta) \biggr] \nonumber \\ && \mbox{} - 4\pi \Biggr]
(m\omega)
- {1 \over 48}
\eta \chi_1 \chi_2 \left[ 247({\bf \hat s_1 \cdot \hat s_2}) - 721({\bf \hat
L_N
\cdot \hat s_1})({\bf \hat L_N \cdot \hat s_2}) \right]
(m\omega)^{4/3} \Biggr\} .
\end{eqnarray}
The evolution of the orbital frequency can be used to calculate the
accumulated orbital phase of the binary.  The orbital phase as observed
by a phase sensitive detector such as LIGO/VIRGO is given
by
\begin{equation}
\Psi \equiv \int^{t_f}_{t_i} \omega dt = \int^{\omega_f}_{\omega_i}
{\omega \over \dot \omega} d\omega ,
\end{equation}
where $t_i$ is the time at which the signal enters the sensitive bandwidth
(corresponding to a lower frequency $\omega_i$ set by seismic noise) and
$t_f$ is the time at which the signal leaves the sensitive bandwidth
(corresponding to an upper frequency $\omega_f$ set by photon shot noise),
the time at which the orbit begins to plunge (corresponding to a
frequency $\omega_f$ of the innermost stable circular orbit), or the
time when the two bodies begin to coalesce.  The result is
\begin{eqnarray}
\Psi  &=& {1 \over 32\eta} \Biggl\{ \left[ (m\omega_i)^{-5/3}
- (m\omega_f)^{-5/3} \right]
+ {5 \over 1008}(743+924\eta) \left[ (m\omega_i)^{-1} - (m\omega_f)^{-1}
\right]
\nonumber \\ && \mbox{}
+ \left[ {5 \over 24} \sum_{i = 1,2} \left[ \chi_i ({\bf \hat L_N
\cdot \hat s_i}) ( 113{m_i^2
\over m^2} +75\eta) \right] - 10\pi \right] \left[
(m\omega_i)^{-2/3} - (m\omega_f)^{-2/3} \right]  \nonumber \\ && \mbox{}
+ {5 \over 48}
\eta \chi_1 \chi_2 \left[ 247({\bf \hat s_1 \cdot \hat s_2})
- 721({\bf \hat L_N
\cdot \hat s_1})({\bf \hat L_N \cdot \hat s_2}) \right] \left[
(m\omega_i)^{-1/3} - (m\omega_f)^{-1/3} \right] \Biggr\} . \label{phase}
\end{eqnarray}
The term with $10\pi$ is the tail contribution to the orbital phase.
Table \ref{phasetable} shows the contribution
of each term to the orbital phase for
several cases.
\narrowtext

\subsection{Radiation-Reaction Effects on the Precessions}

When averaged over a circular orbit, the orbital angular
momentum and spins precess as
\begin{mathletters}
\label{circprec}
\begin{eqnarray}
{\bf \dot L}  &=& {1 \over 2r^3} \left\{ \left[ \left( 4+3{m_2 \over m_1}
\right) {\bf S_1} + \left( 4 + 3
{m_1 \over m_2} \right) {\bf S_2} \right] {\bf \times L_N}
\right. \nonumber \\ && \left. - 3({\bf \hat L_N \cdot S_2})
{\bf S_1 \times \hat L_N}
- 3({\bf \hat L_N \cdot S_1})
{\bf S_2 \times \hat L_N} \right\} , \nonumber \\ && \label{ldot}
\end{eqnarray}
\begin{eqnarray}
{\bf \dot S_1}  = && {1 \over 2r^3} \biggl\{ ({\bf L_N \times S_1})(4+3
{m_2 \over m_1}) + {\bf S_2 \times S_1} \nonumber \\ && \mbox{}
- 3({\bf \hat L_N \cdot S_2})
{\bf \hat L_N \times S_1} \biggr\} ,
\end{eqnarray}
\begin{eqnarray}
{\bf \dot S_2}  = && {1 \over 2r^3} \biggl\{ ({\bf L_N \times S_2})(4+3
{m_1 \over m_2}) + {\bf S_1 \times S_2} \nonumber \\ && \mbox{}
- 3({\bf \hat L_N \cdot S_1})
{\bf \hat L_N \times S_2} \biggr\} .
\end{eqnarray}
Since $\langle {\bf \dot L_{SO}} \rangle = 0$ over an orbit,
the precession of ${\bf L_N}$ is also given by Eq.
(\ref{ldot})
with ${\bf L}$ replaced by ${\bf L_N}$.  Thus we see that for a circular
orbit the magnitudes of the vectors ${\bf L}$, ${\bf L_N}$, and ${\bf S_i}$
are conserved on average during precession in the absence of radiation
reaction.
\end{mathletters}

For a binary system inspiralling due to gravitational radiation, there is
a loss of total angular momentum given by ${\bf \dot J}$.  We assume that
the individual spinning bodies are sufficiently axisymmetric that they
will emit negligible gravitational radiation on their own, so that their
spins are unaffected by radiation damping to the order which we are
considering. (See Ref. \cite{precession}
for a more rigorous argument.)  This means that
the angular momentum loss is entirely from the orbital angular
momentum, and that the total change in orbital angular momentum is
the sum of the precession and the radiation damping.

Let us examine this in more detail for the case of one spinning body,
where we only consider the leading order damping effects.  Then we have
(see Eqs. (\ref{onesdot})-(\ref{spinprec}))
\begin{mathletters}
\label{onespin}
\begin{equation}
{\bf \dot S}  = \omega_p {\bf \hat J \times S} ,
\end{equation}
\begin{equation}
{\bf \dot L_N}  = \omega_p {\bf \hat J \times L_N} -
\epsilon_{RR}{\bf L_N},
\end{equation}
\begin{equation}
{\bf \dot J}   = - \epsilon_{RR}{\bf L_N} ,
\end{equation}
where ${\bf \hat J} = {\bf J}/|{\bf J}|$, and
where $\epsilon_{RR} = (32/5)(\mu/m^2)(m/r)^4$ is the
rate of angular momentum loss
due to gravitational radiation.  Defining $L = |{\bf L_N}|$, $S= |{\bf S}|$,
and $\kappa = \cos^{-1} \left( {\bf \hat L_N \cdot \hat S} \right)$,
\end{mathletters}
\begin{mathletters}
\begin{equation}
J  = |{\bf J}| =  \left\{ L^2 + S^2 + 2LS \cos \kappa \right\}^{1/2},
\end{equation}
\begin{equation}
i  = \cos^{-1} \left( {\bf \hat L_N
\cdot \hat J} \right) =  \cos^{-1} \left\{
\left[ L + S \cos \kappa \right] / J \right\},
\end{equation}
we see that Eqs. (\ref{onespin}) imply that
\end{mathletters}
\begin{mathletters}
\begin{equation}
\dot S  = 0 ,
\end{equation}
\begin{equation}
\dot L  = - \epsilon_{RR} L ,
\end{equation}
\begin{equation}
\dot J  = - \epsilon_{RR} L \cos i ,
\end{equation}
\begin{equation}
\dot \kappa  = 0 ,
\end{equation}
\begin{equation}
{di \over dt}  = {\epsilon_{RR} L \over J} \sin i .
\end{equation}
As the binary inspirals, ${\bf L_N}$ and ${\bf J}$ shrink
and the angle between them grows.  This is shown in Fig.
\ref{simple_precession}.  The angle $\kappa$ between ${\bf L_N}$ and
${\bf S}$ remains fixed as they precess about ${\bf J}$,
so that ${\bf S}$ tips toward ${\bf J}$ and at late
times when ${\bf J} \approx {\bf S}$, $i$ approaches $\kappa$.
Working to first order in $(L\epsilon_{RR})/(J\omega_p)$,
we can show that, if ${\bf J_o}$ is the initial direction of ${\bf J}$,
\end{mathletters}
\begin{equation}
{\bf J} = {\epsilon_{RR} \over \omega_p} \left( {\bf \hat J_o \times L_N}
\right) + \left[ 1 - {\epsilon_{RR} L \over J} (t-t_o) \cos i \right]
{\bf \hat J_o},
\end{equation}
valid for $(t-t_o) << J/(L\epsilon_{RR})$,
which implies that ${\bf J}$ spirals about ${\bf J_o}$ as it shrinks.  As
long as the ratio $\Lambda \equiv (L \epsilon_{RR})/(J \omega_p) << 1$,
then ${\bf \hat J}$
will remain relatively fixed in space, precessing on a tight spiral around its
earlier direction.  If the ratio $\Lambda$ is
not small, however, ${\bf L_N}$ and
${\bf S}$ will still precess about ${\bf J}$, but ${\bf J}$ will start to
tumble. An example of this would be if ${\bf L}$ and ${\bf S}$ were nearly
equal in magnitude, but pointing in opposite directions so that
$J = |{\bf L} + {\bf S}| << L$.  This is the case of ``transitional
precession" described in Ref. \cite{precession}.

If both bodies are spinning, the precessions of ${\bf L}$ and ${\bf S}$
are more complicated (as in
Sec. \ref{sec:orbit}), but the effects of
radiation damping are qualitatively the same.
As the binary inspirals, ${\bf L_N}$ and ${\bf J}$ shrink and the
maximum angle between them grows.  ${\bf J}$ remains relatively
fixed in direction as long as the precession timescale is shorter than the
inspiral timescale, and as long as the ratio $\Lambda$ is small, where we
replace the precession frequency $\omega_p$ with the instantaneous
precession frequencies $\omega_p^{(A)}$.  Fig. \ref{comp_prec_rad} shows
an example of this.  Numerical
evolutions of the precession equations in the presence of radiation-reaction
agree with our description above.

\subsection{Wave Polarization States}

The gravitational radiation emitted by the binary can be written in terms of
its two polarization states $h_+$ and $h_\times$.
The polarization states can be expressed as linear combinations of the
components of $h^{ij}$ in some suitable coordinate system.  Normally one
chooses a coordinate system such that the orbital plane lies in the x-y
plane, and the source and detector are located in
the x-z plane as this simplifies
the equations involved (see Ref. \cite{finn}).
If the bodies are spinning, however, this coordinate system is not fixed
in space.  This does not prevent one from using such a coordinate system
(see Apostolatos {\it et al} \cite{precession}),
but one must remember that it is precessing.
Instead, we choose our coordinate system such that our z-axis
lies along some initial direction of ${\bf J}$.
We define $\Theta = \cos^{-1} \left( {\bf \hat N \cdot \hat J_o} \right)$ and
choose a coordinate system in which ${\bf \hat z} = {\bf \hat J_o}$ and
${\bf \hat N} = \cos \Theta {\bf \hat z} + \sin \Theta {\bf \hat x}$
lies in the x-z plane (see Fig. \ref{source_coordinates}).
(Note that if $\Theta = 0$, ${\bf \hat x}$ can be chosen arbitrarily.)
Following the method of Finn and Chernoff \cite{finn} we define the
radiation coordinate system such that
\begin{mathletters}
\begin{equation}
{\bf e}_z^R = {\bf \hat N} ,
\end{equation}
\begin{equation}
{\bf e}_y^R = {\bf \hat y} = {{\bf \hat J_o \times \hat N} \over \left\{
1 - ({\bf \hat J_o \cdot \hat N})^2 \right\}^{1/2}} ,
\end{equation}
\begin{equation}
{\bf e}_x^R = {\bf \hat y \times \hat N} = {({\bf \hat J_o \cdot \hat N})
{\bf \hat N} - {\bf \hat J_o} \over \left\{
1 - ({\bf \hat J_o \cdot \hat N})^2 \right\}^{1/2}} .
\end{equation}
Then from Eqs. (3.2) and (3.3) of Ref. \cite{finn},
the radiation can be written in terms of its polarization states
\end{mathletters}
\begin{mathletters}
\label{h+hx}
\begin{equation}
h_+ = \case{1}{2} \left\{ \cos^2 \Theta h^{xx} - h^{yy} +
\sin^2 \Theta h^{zz} - \sin 2\Theta h^{xz} \right\},
\end{equation}
\begin{equation}
h_\times = \cos \Theta h^{xy} - \sin \Theta h^{yz}.
\end{equation}
The response of a detector will be the linear combination
\end{mathletters}
\begin{equation}
h = F_+h_+ + F_\times h_\times,
\end{equation}
where $F_+$ and $F_\times$ are the antenna patterns
which depend upon the orientation of the detector with respect to the binary.
\begin{mathletters}
\begin{equation}
F_+ = \case{1}{2} (1 + \cos^2 \theta) \cos 2\phi \cos 2\psi
- \cos \theta \sin 2\phi \sin 2\psi ,
\end{equation}
\begin{equation}
F_\times = \case{1}{2} (1 + \cos^2 \theta) \cos 2\phi \sin 2\psi
+ \cos \theta \sin 2\phi \cos 2\psi ,
\end{equation}
where $(\theta,\phi)$ is the location of the binary with respect to the
detector (whose arms lie along the $\bar x$ and $\bar y$ axes in the
detector's coordinate system, with the $\bar z$ axis in the vertical
direction),
and $\psi$ is the polarization angle between the gravitational waves'
polarization axes and the direction of constant azimuth.
(See Fig. \ref{detector_coordinates}.)  For our definition
of the polarization axes, the polarization angle is given by
\end{mathletters}
\begin{equation}
\psi = \tan^{-1} \left(
{{\bf \hat N \cdot} ( {\bf \hat J_o \times \hat {\bar z}} )
\over  {\bf \hat J_o \cdot \hat {\bar z}}
- ({\bf \hat J_o \cdot \hat N} ) ( {\bf \hat {\bar z} \cdot \hat N)}}
\right). \label{psi}
\end{equation}
If ${\bf \hat N} = \pm {\bf \hat {\bar z}}$ then
\begin{equation}
\psi = \tan^{-1} \left[ -({\bf \hat N \cdot \hat {\bar z}})
{{\bf \hat {\bar y} \cdot \hat J_o} \over
{\bf \hat {\bar x} \cdot \hat J_o} } \right]  .
\end{equation}
If we define $\Phi$ as the orbital phase
with respect to the line of ascending nodes (the point at which the orbit
crosses the x-y plane from below), then for
a circular orbit the polarization
states will be given by
\begin{eqnarray}
h_+ = && {2 \mu \over D} \left( {m \over r} \right)  \biggl\{
Q_+ + P^{0.5}Q_+ \left( {m \over r} \right)^{1/2}
+ PQ_+ \left( {m \over r} \right) \nonumber \\ && \mbox{}
+ P^{1.5}Q_+ \left( {m \over r} \right)^{3/2} \biggr\} ,
\end{eqnarray}
where the quadrupole term is given by
\begin{equation}
Q_+ = - 2 \left[ C_+ \cos 2\Phi + S_+ \sin 2\Phi \right]  , \label{qlist}
\end{equation}
and similarly for $h_\times$ where $+$ is replaced by $\times$, and
where
\begin{mathletters}
\begin{eqnarray}
C_+ =&& \case{1}{2} \cos^2 \Theta \left( \sin^2 \alpha - \cos^2 i \cos^2 \alpha
\right)
+ \case{1}{2} ( \cos^2 i \sin^2 \alpha \nonumber \\ && \mbox{}
- \cos^2 \alpha )
- \case{1}{2} \sin^2 \Theta \sin^2 i - \case{1}{4} \sin 2\Theta \sin 2i
\cos \alpha , \nonumber \\
\end{eqnarray}
\begin{equation}
S_+ =  \case{1}{2} \left( 1 + \cos^2 \Theta \right) \cos i \sin 2\alpha
+ \case{1}{2} \sin 2\Theta \sin i \sin \alpha ,
\end{equation}
\begin{equation}
C_\times =  - \case{1}{2} \cos \Theta \sin 2\alpha \left( 1 + \cos^2 i
\right)
- \case{1}{2} \sin \Theta \sin 2i \sin \alpha ,
\end{equation}
\begin{equation}
S_\times = - \cos \Theta \cos i \cos 2\alpha - \sin \Theta
\sin i \cos \alpha ,
\end{equation}
where $(i,\alpha)$ are the spherical coordinates describing the direction
of ${\bf \hat L_N}$ (see Fig. \ref{source_coordinates}).  We give the
post-Newtonian corrections to $h_+$ and $h_\times$ in Appendix \ref{sec:appc}.
The evolution of $i$ and $alpha$ is given by the precession equations
(\ref{circprec}).  The evolution of $\Phi$ is given by
\end{mathletters}
\begin{equation}
\dot \Phi = \omega - \dot \alpha \cos i .
\end{equation}

Note that our expressions for the polarization states (specifically the
quadrupole terms) are much more complicated than the expressions in
Apostolatos {\it et al} \cite{precession} because we have defined them with
respect to a fixed coordinate system as opposed to a rotating one.  In our
description $\alpha$ varies as $\sim \omega_pt$ as the orbit precesses in
the case of simple precession.  On the
other hand, in our description the polarization angle $\psi$ (\ref{psi}) (and
thus
the antenna patterns
$F_+$ and $F_\times$) is constant during the binary's inspiral, while in
the description of Ref. \cite{precession} it is not.  The two descriptions
are of course equivalent;  we have simply made the complexity of the waveforms
more explicit.

 From the above equations, it can be seen that the signal in the detector
due to the quadrupole waveform
can be written as
\begin{equation}
h = C_Q \cos 2\Phi + S_Q \sin 2\Phi ,
\end{equation}
where
\begin{mathletters}
\begin{equation}
C_Q = - {4 \mu \over D} \left( {m \over r} \right) \left[
C_+F_+ + C_\times F_\times \right] ,
\end{equation}
\begin{equation}
S_Q = - {4 \mu \over D} \left( {m \over r} \right) \left[
S_+F_+ + S_\times F_\times \right] .
\end{equation}
The signal can be rewritten as
\end{mathletters}
\begin{equation}
h =  A_Q \cos \left[ 2\Phi - \delta_Q \right] , \label{qsig}
\end{equation}
where
\begin{mathletters}
\begin{equation}
A_Q = \left\{ C_Q^2 + S_Q^2 \right\}^{1/2} , \label{quadamp}
\end{equation}
\begin{equation}
\delta_Q = \tan^{-1} \left( S_Q/C_Q \right) .
\end{equation}
\end{mathletters}

\subsection{Small Inclination Angles}

Let us examine the waveform in the limit of a small inclination angle $i$.
This limit will be valid if the total spin ${\bf S}$ is nearly aligned
with the orbital angular momentum ${\bf L}$ or if $|{\bf S}| << |{\bf L}|$.
In Fig. \ref{small_inc} we show the region for which the precession angle $i$
is less than $0.2$ for an equal mass system.
If we expand Eq. (\ref{qlist}) through ${\cal O}(i^2)$,
the overall signal amplitude due to the quadrupole term (\ref{quadamp})
will be given by
\widetext
\begin{eqnarray}
A_Q = && {2 \mu \over D} \left( {m \over r} \right) \biggl\{
F_+^2(\theta,\phi,\psi) \biggl[ \left( 1 + \cos^2 \Theta \right)^2
+ 2i\sin 2\Theta \left( 1 + \cos^2 \Theta \right)
\cos \alpha - i^2 \biggl( 1 - 2\cos^2 \Theta
\nonumber \\ && \mbox{} + 5\cos^4 \Theta \biggr)
+ 3 i^2 \sin^2 \Theta \left( 1 + \cos^2 \Theta \right) \cos 2\alpha \biggr]
+ 4F_\times^2(\theta,\phi,\psi) \Bigl[ \cos^2 \Theta
+ i \sin 2\Theta \cos \alpha \nonumber \\ && \mbox{}
- i^2 \cos 2\Theta \Bigr] + F_+(\theta,\phi,\psi)F_\times(\theta,\phi,\psi)
\left[ - 2 i \sin^3 \Theta \sin \alpha + 3i^2 \sin^2 \Theta \cos \Theta
\sin 2\alpha \right] \biggr\}^{1/2}. \nonumber \\ \label{smalli}
\end{eqnarray}
Notice that to lowest order in $i$, the amplitude is constant, independent
of the precession angle $\alpha$, and that the
modulations are of ${\cal O}(i)$ and have the same frequency as the precession
frequency, $d\alpha/dt$.  For some detector
orientations (see below), the ${\cal O}(i)$ terms
are suppressed and the modulations will
be of ${\cal O}(i^2)$ and go as twice the
precession frequency.
\narrowtext

We can use Eq. (\ref{smalli}) to explain the features of
Fig. \ref{amplitude_modulations}.
Note that this case has been presented in Cutler {\it et al} \cite{jugger},
and in Apostolatos {\it et al} \cite{precession} for a single
detector orientation which corresponds to $\gamma = i$.  In
Fig. \ref{amplitude_modulations}, $\Theta = \pi/2$,
$F_+ = \cos 2\gamma$, $F_\times = \sin 2\gamma$,
and initially $i \approx 0.1$, so that
\begin{eqnarray}
A_Q = && {2 \mu \over D} {m \over r} \Bigl\{ \cos^2 2\gamma \left[ 1 - i^2
+ 3i^2 \cos 2\alpha \right] \nonumber \\ && \mbox{}
- 2i \sin 4\gamma \sin \alpha
+ 4 i^2 \sin^2 2\gamma \Bigr\}^{1/2} ,
\end{eqnarray}
For $\gamma = 0$, the modulations of the amplitude will have a frequency of
twice the precession frequency and will be only a few percent
($\approx (3/2)i^2$) of
the overall amplitude.  For $\gamma = \pi/4$, $A_Q \approx 4i (\mu/D)(m/r)$
which is roughly $20\%$ (initially) of the unmodulated
amplitude in the previous case and is unmodulated
through ${\cal O}(i^2)$.  The other three cases are
modulated by a term with a frequency
of twice the precession frequency, and another term with the precession
frequency.  For $\gamma = \pi/8$, the
dominant modulation is of ${\cal O}(i)$ and
varies with the precession frequency.  In the other two cases
($\gamma = i/2$ and $\gamma = i$), both
modulations are of ${\cal O}(i^2)$ and
contain both $\alpha$ and $2\alpha$ terms,
which leads to their forms in
Fig. \ref{amplitude_modulations}.  Finally, let $\gamma = i + \pi/4$.  Then
$A_Q \approx 8i (\mu/D)(m/r) | \cos (\alpha/2 - \pi/4) |$ which
corresponds with Fig. 6 of
Ref. \cite{precession}.

\section{Results for Specific Systems}
\label{sec:results}

In this section we will describe several specific cases involving a variety
of masses and spins.  For some cases, we have solved the equations of motion
and precession numerically and used them to calculate the emitted
gravitational waveform.  In all the examples we will assume the binary's
orbit has been circularized prior to entering the frequency bandwidth
of a LIGO-type detector.

\subsection{Nonspinning Bodies}

The case in which neither body is spinning has been studied in
previous papers \cite{linc,agw92,isco,agwtail}.  Here we only wish
to mention that if neither body is spinning, the orbital plane
remains fixed in space, and the waveform's amplitude increases
monotonically (apart from post-Newtonian modulations on orbital
timescales, see Fig. \ref{gravitational_waveform}) as the binary inspirals.

\subsection{Spins Perpendicular to the Orbital Plane}

If the spins of the bodies are aligned with the orbital angular momentum,
the system evolves in a manner qualitatively similar to the case of
nonspinning bodies.  Since the spins and orbital angular momentum are
aligned, none of them precesses so that the orbital plane remains
fixed in space.  The only effects the spins will have are a contribution
to the orbital phase, and a correction to the amplitude of the waveform.

The contribution to the orbital phase will be important as it affects
the accumulated phase of the waveform.  Since matched templates will be used
to obtain information about the binary from the waveform, any effect which
causes the phase to change by one cycle over the thousands in the bandwidth
of the detector will be important.  If the spin contribution to the phase
can be separated from the other contributions, it could be used to make an
accurate determination of the spins.  However, preliminary
studies by Cutler and
Flanagan \cite{flanagan} suggest that the spin contribution cannot be
separated cleanly from other contributions, but instead are strongly
correlated with them,
thus making the determination of the individual masses
and spins more difficult.  Table \ref{phasetable}
compares the spin contribution to
the orbital phase with other contributions for various binary systems.

The leading-order correction to the waveform due to the spins is a full
post-Newtonian order higher than the quadrupole part of the waveform.  Thus
it will be small until the late stages of the binary's inspiral.
In Fig. \ref{gravitational_waveform}
we compare the spin contributions to the waveform with other contributions.

\subsection{One Spinning Body}

In the cases in which only one of the bodies is spinning,
our numerical results agree with our
analytic description of Sec. \ref{sec:circ}.
As the binary inspirals, ${\bf L}$ and ${\bf S}$
precess about ${\bf J}$, with both ${\bf L}$ and ${\bf J}$ shrinking, and ${\bf
L}$
tipping away from ${\bf J}$ while ${\bf S}$ tips toward ${\bf J}$
(see Fig. \ref{simple_precession}).  The angle
between ${\bf L}$ and ${\bf S}$ remains constant.
${\bf J}$ remains
relatively fixed in direction
unless ${\bf L}$ and ${\bf S}$
are nearly antialigned and equal in magnitude, in which case the ``transitional
precession" described by Apostolatos {\it et al} \cite{precession} occurs.

\subsubsection{Dependency of modulation on detector orientation and location}

The precession of the orbital plane causes the waveform amplitude to be
modulated since the orientation between the orbital plane and the detector
is changing.  The form of this modulation depends on the orientation of the
detector and its location with respect to the source.  In
Fig. \ref{amplitude_modulations}, which was generated from numerical solutions,
we show how the modulation changes for different orientations of a detector at
a fixed location.
Notice that the size and shape of the modulations
vary greatly.
These modulations are discussed qualitatively in Sec. \ref{sec:circ}.

\subsubsection{Effects of higher order parts of the waveform}

So far, we have just examined how the precession of the orbital plane
modulates the dominant quadrupole part of the waveform.  The precession
will also modulate the amplitude of the post-Newtonian corrections
to the waveform.  This is illustrated in Fig. \ref{high-order_modulations}.
Recall that the
different contributions to the waveform have different dependences
on the orbital phase, so that the overall waveform amplitude is not the
sum of the individual higher-order amplitudes.
(Note that this is true even without spins.)

\subsection{Two Spinning Bodies}

We are not able to solve the general two-spinning-body problem analytically.
We therefore present numerical solutions of the equations of motion and
precession.  Most cases involving two spinning bodies are qualitatively
similar to the case of one spinning body.  The main difference lies in the
fact that the total spin ${\bf S}$ is not constant as the spins precess,
but rather oscillates between some ${\bf S_{min}}$ and ${\bf S_{max}}$.
This causes the angle $i$ between ${\bf L}$ and ${\bf J}$ to oscillate
as ${\bf L}$ precesses about ${\bf J}$ in addition to its overall increase
due to gravitational radiation damping.  In most cases where spin effects
will be important, these oscillations will be small.
In each of the following examples, we will examine a pair of inspirals which
have the same initial orientation of ${\bf L}$ and ${\bf S}$,
but in one case only one of the bodies is spinning, while in the other
case both bodies are spinning.

If one of the spins
is much smaller than the other, it can be viewed as a perturbation of the
simple precession involving the larger spin and the orbital angular momentum.
Recall that the spin ratio will go as $|{\bf S_1}|/|{\bf S_2}| =
(\chi_1/\chi_2)(m_1/m_2)^2$, so that if $m_1 << m_2$ and $\chi_2$ is not
small, then the spin of the smaller body will be much smaller than that of
the larger body, and the above argument will hold.
The overall shape of the modulations of the waveform amplitude
is the same in both the one spin case and the two spin case.  The main
difference is that the precession frequency is slightly different in the
two cases, which could lead to a noticeable effect on the phase of the
waveform.  Fig. \ref{large_mass-ratio} shows
the difference
in precession rates.

If the masses of the two bodies are equal, Eq. (\ref{sdot}) implies that
the only change in the evolution of ${\bf S}$ (and thus in $|{\bf S}|$)
compared to the one-spin case will be from the spin-spin coupling.
This coupling is weak so that it can be viewed as a perturbation on the
simple precession case.  Fig. \ref{equal_mass} illustrates this case.
Notice that
while the individual spins precess wildly and in effect exchange places,
the total spin and orbital angular momentum remain approximately fixed
relative to each other ($\kappa \approx const.$).  This leads to
the overall shape of the modulations of the waveform amplitude
to be very similar for the one-spin case and the two-spin case.  As in the
previous example there is a
difference in the precession frequency between the
two cases, which could lead to a noticeable effect on the phase of the
waveform.

In some cases, however, the oscillations in $i$ can be significant, and
cannot be viewed as a perturbation of a simple precession case.
Fig. \ref{wild_precession}
illustrates this case, for equal spins but different masses.
Even though the spins are equal in magnitude, the
different masses will cause them to precess at different rates (see
Eq. (\ref{sdot})).  At some
times the two spins will align and $i$ will be at its largest value.  At
other times the two spins will almost cancel each other, and ${\bf L}$
will be aligned with ${\bf J}$.  Thus the orbital plane will tilt back and
forth as it precesses about ${\bf J}$.  This leads to substantial differences
between the two spin case and the one spin case, as illustrated by the
complex modulations in Fig. \ref{wild_precession}.

Another case in which the second spin will be important is that of
``transitional precession" described
by Apostolatos {\it et al} \cite{precession}.
Recall that in this case ${\bf L}$ and ${\bf S}$ are almost cancel each
other, so that any perturbations in ${\bf S}$ will cause noticable effects.
See Ref. \cite{precession} for more details.

\section{Conclusions}
\label{sec:conclusion}

In this paper we have examined the importance of spin-orbit and spin-spin
effects on the inspiral of a coalescing binary system of
compact objects and on the gravitational
radiation emitted from such a system.  The inclusion of spin effects makes
the study of coalescing binary systems much more complicated because of
the extra degrees of freedom for the orbit, and the extra parameters upon which
the waveform will depend.  On the other hand, if the effects due to the spins
can be separated from other post-Newtonian effects, more information
about the binary system can be extracted from the observed
waveforms.

The spins of the bodies have two major effects on the inspiral of the binary.
As long as the spins are not perpendicular to the orbital plane, the orbital
plane will precess, thus changing its orientation in space.  In most cases,
this precession will be a relatively simple precession of the orbital
angular momentum ${\bf L}$ about the total angular momentum
${\bf J}$.  In some cases, however, the precession can be quite complicated
(See Sec. \ref{sec:orbit}).
In addition to the precession effect, the spins will contribute to the
evolution of the orbital phase, in the same manner that other post-Newtonian
terms contribute (See Sec. \ref{sec:circ} and Table \ref{phasetable}).
The spin-orbit contribution can be of
the same magnitude as the post-Newtonian and leading-order tail contributions
to the orbital phase.  The spin-spin contribution, on the other hand, is
quite small.  These contributions to the orbital phase will change the rate
of inspiral.

The spins will change the amplitude of the observed waveform in several ways.
The major effect the spins have is due to the precession of the orbital plane.
This causes the orientation of the orbit to be changed with respect to the
detector, which results in modulation of the amplitude of the waveform (See
Sec. \ref{sec:circ}, and Fig. \ref{amplitude_modulations}).
The size and shape of the modulations are sensitive to
the location of the detector with respect to the binary system and to the
orientation of the detector arms.  The spins also directly contribute
to the amplitude of the waveform in the same manner that other higher
post-Newtonian terms contribute (See Sec. \ref{sec:rad} and Fig.
\ref{gravitational_waveform}).  The direct spin contribution
to the amplitude may be difficult to detect because it
is much smaller than the quadrupole term and the leading-order
post-Newtonian terms until very late in the inspiral.
The modulations due to precession, however, will be quite
noticeable in many cases involving spinning bodies.

The spins will also affect the phase of the observed waveform in several ways.
Since the spins affect the evolution of the orbital phase, this will in turn
affect the evolution of the phase of the gravitational waveform since it is
related to the orbital phase.  The phase will also be affected by the
precession of the orbital plane since the point from which the orbital
phase is measured is itself moving (See Sec. \ref{sec:circ}).
These two effects are independent of one
another as one may be present without the other.  These effects are very
important, however, as the sensitivity with which the phase of the waveform
can be measured is currently thought to be the best way of extracting
information about the binary system.

In general, the effects of the spins on the waveform amplitude
will be small for the case of two
coalescing neutron stars. One reason is
that for the majority of the time in which the frequency
of the gravitational waves from the
binary are in the bandwidth of a LIGO-type detector, the orbital angular
momentum will be much larger than the spin angular momentum.
For example, over $95\%$ of the gravitational wave cycles will occur
between $r=174m$ and $r=37m$ for two $1.4 M_\odot$ neutron stars, and as
Fig. \ref{momentum_size} shows, ${\bf L}$ is at
least $5$ times larger than ${\bf S}$.
This means that the inclination angle $i$ will be small throughout most, if not
all, of the observed inspiral.  This implies that the small inclination angle
approximation will hold, so that the amplitude modulations will be on the order
of $i$.  For small enough $i$, one might be able to ignore the modulations to
first order and
treat the orbital plane to be fixed perpendicular to ${\bf J}$ (which is
relatively fixed) instead of ${\bf L}$ (which is precessing).  Then the
modulations can be treated as perturbations of ${\cal O}(i)$.

While the modulations of the amplitude of the waveform may be small,
the spin effects on the phase of the waveform are significant for
a coalescing binary system of neutron stars.  While the spin-orbit
contribution to the orbital phase is less than one percent of the total
(see Table \ref{phasetable}), a neutron star
with $\chi = S_{NS}/m_{NS}^2$ as small as $0.01$
can cause the
accumulated gravitational wave phase to change by a cycle from an
equivalent system with no spins over the bandwidth of 10 Hz to 1000 Hz.
A further contribution to the
waveform's phase will come from the precession of the orbital plane.
This contribution should vary at the precession frequency.  Since
${\bf L} >> {\bf S}$ for a binary neutron star coalescence, we may
substitute ${\bf L}$ for ${\bf J}$ in Eq. (\ref{spinprec}) and integrate
over the inspiral, finding that the binary precesses roughly
60-70 times over the detector's bandwidth of 10 Hz to 1000 Hz.

Spin effects can be very important for coalescing binary systems with a
very massive black hole and a
neutron star.  Since the spins can dominate the orbital
angular momentum, large precession angles are possible, leading to very
large amplitude modulations.  Furthermore, the spin-orbit contribution
to the orbital phase is a larger percentage of the total.

We have examined how the spins of the body affect the inspiral of the binary
system and the gravitational radiation emitted therefrom.  There are many
interesting questions that remain to be examined.  One is the inverse
problem:  given an observed waveform, how much information about the spins
of the bodies can be extracted.  A related question is how much will
spin effects complicate the extraction of the masses and other information
from the waveform.  Preliminary studies by Cutler and Flanagan \cite{flanagan}
indicate that the spin-orbit contribution to the orbital phase will
be difficult to separate from the post-Newtonian contribution which
is used to determine the mass ratio because of their similar dependences on
$(m/r)$ or $(m\omega)$ in the evolution of the phase (see Eq. \ref{phase}).
Further studies are needed to
determine whether Newtonian templates without explicit spin contributions
to the waveforms can be used on
systems which have spins without significant loss of signal-to-noise
ratio.  If not, then the number of templates which will be needed
to study coalescing binary systems of compact objects will
increase significantly.

\acknowledgements

We would like to thank Clifford M. Will for his many helpful suggestions.
We are grateful to Kip Thorne for sharing his early calculations of spin
effects with us and for encouraging us to pursue the problem.  We would like
to thank Alan G. Wiseman for his contribution to our early work on spin
effects.  We also acknowledge useful discussions with Theocharis Apostolatos,
Curt Cutler, Sam Finn, Bala Iyer, Craig Lincoln, and Eric Poisson.  This work
was
supported in part by National Science Foundation Grants No. 92-22902 and
93-18152 (ARPA supplemented), and by
NASA Grant No. NAGW 3874.

\appendix

\section{Spin Supplementary Conditions}
\label{sec:appA}

When examining systems containing spinning bodies, it is important to
note that the form of the spin-orbit acceleration ${\bf a}_{SO}$ is
not unique, but rather depends on a ``spin supplementary condition"
(SSC).  For example, three different forms of ${\bf a}_{SO}$
\begin{mathletters}
\begin{eqnarray}
{\bf a}^{(I)}_{SO} =&& {1 \over r^3}
\biggl\{ 6 {\bf \hat n} [( {\bf \hat n} \times
{\bf v} ) {\bf \cdot} (2{\bf S} + {\delta m \over m}{\bf \Delta} )]
\nonumber \\ && \mbox{} - [ {\bf v} \times (7
{\bf S}+3{\delta m \over m}{\bf \Delta})]
\nonumber \\ && \mbox{} + 3 \dot r [ {\bf \hat n} \times
(3{\bf S} + {\delta m \over m}{\bf \Delta} )] \biggr\} ,
\end{eqnarray}
\begin{eqnarray}
{\bf a}^{(II)}_{SO}  =&& {1 \over r^3} \biggl\{ {3 \over 2}
{\bf \hat n} [( {\bf \hat n} \times
{\bf v} ) {\bf \cdot} (7{\bf S} + 3{\delta m \over m}{\bf \Delta} )]
\nonumber \\ && \mbox{} - [ {\bf v} \times (7
{\bf S}+3{\delta m \over m}{\bf \Delta})]
\nonumber \\ && \mbox{} + {3 \over 2}  \dot r [ {\bf \hat n} \times
(7{\bf S} + 3{\delta m \over m}{\bf \Delta} )] \biggr\} ,
\end{eqnarray}
\begin{eqnarray}
{\bf a}^{(III)}_{SO}  =&& {1 \over r^3} \biggl\{ 3 {\bf \hat n}
[( {\bf \hat n} \times
{\bf v} ) {\bf \cdot} (3{\bf S} + {\delta m \over m}{\bf \Delta} )]
\nonumber \\ && \mbox{} - [ {\bf v} \times (7
{\bf S}+3{\delta m \over m}{\bf \Delta})]
\nonumber \\ && \mbox{} + 6 \dot r [ {\bf \hat n} \times
(2{\bf S} + {\delta m \over m}{\bf \Delta} )] \biggr\} ,
\end{eqnarray}
are given by the three different SSC's
\end{mathletters}
\begin{mathletters}
\begin{equation}
S_A^{\mu\nu} {u_A}_{\nu} = 0 , \label{ssc1}
\end{equation}
\begin{equation}
2{S_A}_{i0} + {S_A}_{ij}v_A^j = 0 , \label{ssc2}
\end{equation}
\begin{equation}
S_A^{i0} = 0 ,
\end{equation}
respectively, where
$u_A^\mu$ is the four-velocity of the center-of-mass world line
$X_A^\mu$ of body $A$, and
\end{mathletters}
\begin{equation}
S_A^{\mu\nu} \equiv 2 \int_A (x^{[\mu} - {X_A}^{[\mu}) \tau^{\nu ] 0}
 d^3x , \label{spindef}
\end{equation}
where $\tau^{\mu\nu}$ denotes the stress-energy tensor of matter plus
gravitational fields satisfying ${\tau^{\mu\nu}}_{,\nu}=0$, and
square brackets around indices denote antisymmetrization.
Note that the spin vector $\bf S$
of each body is defined by
$S_A^i = {1 \over 2} \epsilon_{ijk}S_A^{jk}$.

Barker and O'Connell \cite{sscshift} showed that these different
forms of ${\bf a}_{SO}$ are equivalent if one takes into account
that the different SSC's are related to different locations ${\bf X_A}$ of the
center of mass of each body.  Furthermore they found transformations
from the center-of-mass definition given by one SSC to that of
another.  For example
\begin{equation}
{X_A^i}^{(II)} \longrightarrow {X_A^i}^{(I)} + {1 \over 2m_A}
\left( {\bf v_A \times S_A}
  \right)^i , \label{rshift}
\end{equation}
This shift in the center-of-mass world line is of
post-Newtonian order, so it can be neglected
at lowest order.

In Sec. \ref{sec:orbit}, we chose to use
the form of ${\bf a}_{SO}$ given by the first SSC
(\ref{ssc1}), since it is covariant.  In doing so, we have
chosen a center-of-mass definition through the SSC, and we must
insure that any future center-of-mass definitions we use are
consistent with this choice.  In Sec. \ref{sec:rad},
we use an integral definition
of the center of mass of each body,
\begin{equation}
 x_A^i = {1 \over m_A} \int\limits_A x^i \rho^* \left( {\bf x} \right)
  \left[ 1 + {1 \over 2} \bar v_A^2 + \Pi - {1 \over 2} \bar U_A \right]
  d^3x  ,
\label{intcm}
\end{equation}
to evaluate the BDI multipoles, where
\begin{equation}
m_A = \int\limits_A \rho^* \left( {\bf x} \right) \left[ 1 + {1 \over 2}
\bar v_A^2 + \Pi - {1 \over 2} \bar U_A \right] d^3x ,
\end{equation}
where $ \bar v_A^i = v^i - v_A^i$, $v_A^i = dx_A^i /dt$,
and $ \bar U_A $ is the Newtonian potential produced by the $A$-th body
itself.
This is a ``natural" post-Newtonian definition since it uses the total
mass, kinetic energy, internal energy, and gravitational potential
energy of each body as the weighting factor.  However,
it turns out that this definition of the
center of mass is not related to the first SSC which we chose for our
equations of motion, but rather to the second SSC (\ref{ssc2}).
This can be seen by the following argument.

It is straightforward to show that the three SSC's can be rewritten
in the form
\begin{equation}
S_A^{i0} - k S_A^{ij} v_A^j = 0 , \label{kssc}
\end{equation}
where $k=1$ for the first SSC, $k=1/2$ for the second, and $k=0$ for the
third.
 From Eq. (\ref{spindef}) we see that
\begin{equation}
S_A^{i0} =  \int_A (x^i - X_A^i) \tau^{00}
 d^3x , \label{si0}
\end{equation}
since the integration is done at constant time.
It is straightforward to show that, to post-Newtonian order,
\begin{equation}
\tau^{00} = \left( 1 + 4U \right) T^{00} - {7 \over 8\pi}
| {\bf \nabla} U |^2, \label{tau}
\end{equation}
Substituting Eq. (\ref{tau}) into Eq. (\ref{si0}) and integrating the
$|{\bf \nabla} U |^2$ terms by parts, we obtain
\begin{equation}
S_A^{i0} =  \int_A \rho^*({\bf x}) (x^i - X_A^i) \left[ 1 +
{1 \over 2} v^2 - {1 \over 2} U + \Pi \right] d^3x .
\end{equation}
Using our integral definition of the mass and center of mass of body $A$
(\ref{intcm}), we obtain
\begin{eqnarray}
S_A^{i0} &=& m_A x_A^i \left( 1 + {1 \over 2} v^2 - {1 \over 2} \sum_B
{m_B \over r_{AB}} \right) + {1 \over 2} v_A^j S_A^{ij} \nonumber \\ &&
- m_A  \left( 1 + {1 \over 2} v^2 - {1 \over 2} \sum_B
{m_B \over r_{AB}} \right) X_A^i . \label{saio}
\end{eqnarray}
Thus our integral definition of the center of mass corresponds to the
SSC definition ($x_A^i = X_A^i$), if $S^{i0} = {1 \over 2}v_A^jS^{ij}$,
i.e. the second SSC.  Imposing the general SSC (\ref{kssc}) on
Eq. (\ref{saio}), we see that, to post-Newtonian order,
\begin{equation}
x_A^i =  {X_A^i}^{(k)} + {2k-1 \over 2m_A} S_A^{ij} v_A^j ,
\end{equation}
Thus in order that the center of mass used in the STF moments be consistent
with that used in the equations of motion, we must transform from
$x_A^i = X_A^{i(II)}$ to $X_A^{i(I)}$ in the moments by using
Eq. (\ref{rshift}) (See Sec. \ref{sec:rad}).

This is also consistent with Brumberg \cite{brumberg} who derived the
equations of motion from our integral definition of the center of mass,
and found the form of ${\bf a}_{SO}$ to be that of the second SSC
without ever having mentioned the concept of an SSC.

Another check of the consistency of our argument is seen by evaluating
the BDI mass dipole moment \cite{bd,di}, given by
\begin{equation}
I^i =   \sum_A \left\{ \int_A x^i \rho^*({\bf x}) \left[
1 + {1 \over 2} v^2 - {1 \over 2} U + \Pi \right] d^3x \right\}.
\end{equation}
Evaluating the integral in the manner used to evaluate the mass quadrupole
(in Sec. \ref{sec:rad}), we obtain
\begin{eqnarray}
I^i = && \sum_A \Biggl\{ m_Ax_A^i \left[ 1 + {1 \over 2} v_A^2
- {1 \over 2} \sum_{B \neq A} {m_B \over r_{AB}} \right] \nonumber \\
&& \mbox{} + {1 \over 2}  \left({\bf v_A
\times S_A} \right)^i \Biggr\},
\end{eqnarray}
where we have dropped the post-Newtonian terms as they do not affect the
calculation.  We then use Eq. (\ref{rshift}) to shift the center of mass
$x_A^i$ from that defined by the second SSC $X_A^{i(II)}$
to that of our equations of motion, $X_A^{i(I)}$.
The result is
\begin{eqnarray}
I^i = && \sum_A \Biggl\{ m_Ax_A^i \left[ 1 + {1 \over 2} v_A^2
- {1 \over 2} \sum_{B \neq A} {m_B \over r_{AB}} \right] \nonumber \\
&& \mbox{} +  \left({\bf v_A
\times S_A} \right)^i \Biggr\}, \label{massdipole}
\end{eqnarray}
where $x_A^i$ now refers to $X_A^{i(I)}$.
Blanchet and Damour \cite{bd} have proved on general grounds that
\begin{equation}
{d^2 \over dt^2} I_i = 0. \label{vanish}
\end{equation}
Taking two time derivatives of Eq. (\ref{massdipole}) and substituting
our equations of motion where appropriate, we verify that Eq. (\ref{vanish})
does hold.  (Alternatively, we could have taken two time derivatives
of the mass dipole before shifting the center of mass of each body and
seen that ${\bf a}^{II}_{SO}$ is needed to insure that Eq. (\ref{vanish})
holds.)

\section{Post-Newtonian corrections to the waveform for a circular orbit}
\label{sec:appc}

If we define $\Phi$ as the orbital phase
with respect to the line of ascending nodes (the point at which the orbit
crosses the x-y plane from below), then for
a circular orbit the polarization
states will be given by
\begin{equation}
h_+ = {2 \mu \over D} \left( {m \over r} \right)  \left\{
Q_+ + P^{0.5}Q_+ \left( {m \over r} \right)^{1/2}
+ PQ_+ \left( {m \over r} \right) \right\} ,
\end{equation}
where, for simplicity we include only the post-Newtonian terms
through $PQ_+$, and
\widetext
\begin{mathletters}
\label{hpluslist}
\begin{equation}
Q_+ = - 2 \left[ C_+ \cos 2\Phi + S_+ \sin 2\Phi \right]  ,
\end{equation}
\begin{eqnarray}
P^{0.5}Q_+ =&& {1 \over 4} {\delta m \over m} \bigl[
9 \left( aS_+ + bC_+ \right) \cos 3\Phi
+ 9 \left( bS_+ - aC_+ \right) \sin 3\Phi \nonumber \\ && \mbox{}
+  \left( 3aS_+ - 3bC_+ -2bK_+ \right) \cos \Phi
-  \left( 3bS_+ + 3aC_+ -2aK_+ \right) \sin \Phi \bigr] ,
\end{eqnarray}
\begin{eqnarray}
PQ_+ =&& {8 \over 3} (1-3\eta) \left\{
 \left[ (a^2-b^2)C_+ - 2abS_+ \right] \cos 4\Phi
+ \left[ (a^2-b^2)S_+ + 2abC_+ \right] \sin 4\Phi \right\} \nonumber \\ &&
+ DC_+ \cos \Phi + DS_+ \sin \Phi
+  {1 \over 6} \Bigl\{ \left[ 4(1-3\eta)(a^2+b^2) + (19-3\eta) \right] Q_+
\nonumber \\ && \mbox{}
- 4(1-3\eta) \left[ (a^2-b^2) \cos 2\Phi + 2ab \sin 2\Phi \right] K_+
\Bigr\} ,
\end{eqnarray}
and similarly for $h_\times$ where $+$ is replaced by $\times$, and
where
\end{mathletters}
\narrowtext
\begin{mathletters}
\begin{eqnarray}
C_+ =&& \case{1}{2} \cos^2 \Theta \left( \sin^2 \alpha - \cos^2 i \cos^2 \alpha
\right)
+ \case{1}{2} ( \cos^2 i \sin^2 \alpha
\nonumber  \\ && - \cos^2 \alpha )
- \case{1}{2} \sin^2 \Theta \sin^2 i - \case{1}{4} \sin 2\Theta \sin 2i
\cos \alpha , \nonumber \\
\end{eqnarray}
\begin{equation}
S_+ =  \case{1}{2} \left( 1 + \cos^2 \Theta \right) \cos i \sin 2\alpha
+ \case{1}{2} \sin 2\Theta \sin i \sin \alpha ,
\end{equation}
\begin{eqnarray}
K_+ = && \case{1}{2} \cos^2 \Theta \left( \sin^2 \alpha + \cos^2 i \cos^2
\alpha
\right)
- \case{1}{2} (\cos^2 i \sin^2 \alpha
\nonumber \\ && + \cos^2 \alpha
+ \case{1}{2} \sin^2 \Theta \sin^2 i + \case{1}{4} \sin 2\Theta \sin 2i
\cos \alpha , \nonumber \\
\end{eqnarray}
\begin{equation}
DC_+ =  - {1 \over m^2} \left[ \Delta^y \sin \alpha \cos \Theta
+ d \cos \alpha \right] ,
\end{equation}
\begin{equation}
DS_+ =  - {1 \over m^2} \left[ c \Delta^y  - d \cos i \sin \alpha \right] ,
\end{equation}
\begin{equation}
C_\times =  - \case{1}{2} \cos \Theta \sin 2\alpha \left( 1 + \cos^2 i
\right)
- \case{1}{2} \sin \Theta \sin 2i \sin \alpha ,
\end{equation}
\begin{equation}
S_\times = - \cos \Theta \cos i \cos 2\alpha - \sin \Theta
\sin i \cos \alpha ,
\end{equation}
\begin{equation}
K_\times = - \case{1}{2} \cos \Theta \sin 2\alpha \sin^2 i
+ \case{1}{2} \sin \Theta \sin 2i \sin \alpha ,
\end{equation}
\begin{equation}
DC_\times = {1 \over m^2} \left[ \Delta^y \cos \alpha
- d \cos \Theta \sin \alpha \right] ,
\end{equation}
\begin{equation}
DS_\times = {1 \over m^2} \left[ - \Delta^y \cos i \sin \alpha
+ cd  \right] ,
\end{equation}
\begin{equation}
a = - \sin \Theta \sin \alpha ,
\end{equation}
\begin{equation}
b = \cos \Theta \sin i - \sin \Theta \cos i \cos \alpha ,
\end{equation}
\begin{equation}
c = \cos \Theta \cos i \cos \alpha + \sin i \sin \Theta ,
\end{equation}
\begin{equation}
d = \Delta^z \sin \Theta - \Delta^x \cos \Theta ,
\end{equation}
where $(i,\alpha)$ are the spherical coordinates describing the direction
of ${\bf \hat L_N}$ (see Fig. \ref{source_coordinates}).
\end{mathletters}

 From the above equations, it can be seen that the signal in the detector
can be written as a series
\begin{equation}
h = \sum_{n=1}^{2(N+1)} \left\{ C_n \cos n\Phi + S_n \sin n\Phi \right\} ,
\end{equation}
which can be rewritten as
\begin{equation}
h = \sum_{n=1}^{2(N+1)}  A_n \cos \left[ n\Phi - \delta_n \right] , \label{sig}
\end{equation}
where
\begin{mathletters}
\begin{equation}
A_n = \left\{ C_n^2 + S_n^2 \right\}^{1/2} , \label{amp}
\end{equation}
\begin{equation}
\delta_n = \tan^{-1} \left( S_n/C_n \right) ,
\end{equation}
and $N$ is the final post-Newtonian order beyond the quadrupole term to which
the waveform is calculated.
For example, if
we look at only the quadrupole contribution to the waveform, then
only the $n=2$ terms will be nonzero and
\end{mathletters}
\begin{mathletters}
\label{qamp}
\begin{equation}
C_2 = - {4 \mu \over D} \left( {m \over r} \right) \left[
C_+F_+ + C_\times F_\times \right] ,
\end{equation}
\begin{equation}
S_2 = - {4 \mu \over D} \left( {m \over r} \right) \left[
S_+F_+ + S_\times F_\times \right] .
\end{equation}
\end{mathletters}

 From Eqs. (\ref{hpluslist}) we see that the post-Newtonian corrections
to the waveform depend on different
harmonics of the orbital frequency than the quadrupole term.  Eq. (\ref{sig})
splits the signal into the terms depending on the different harmonics.  Each
term will have its amplitude modulated by the precession of the orbital plane.

\section{A comparison with test mass calculations}
\label{sec:appB}

Recently, Poisson \cite{poisson} has calculated the gravitational waveform
and energy loss due to gravitational radiation from a particle in a circular,
equatorial orbit around a slowly rotating black hole.  Poisson's calculations
were done using the Teukolsky perturbation formalism, which is completely
different from the post-Newtonian calculations we used.  In this appendix, we
show that our results for the waveform and energy loss agree with those of
Poisson in the appropriate limit.  Since we are only interested in the
spin-orbit terms, we will neglect all the other post-Newtonian terms.

Poisson gives the energy loss as
\begin{equation}
{dE \over dt} = - {32 \over 5} \left( {\mu \over m} \right)^2 (m\omega)^{10/3}
\left[ 1 - {11 \over 4} \chi_{BH} (m\omega) \right]
\end{equation}
and the spin-orbit contributions to the waveform as
\begin{mathletters}
\begin{eqnarray}
h_+  = && {2\mu \over D} (m\omega)^{2/3} \biggl\{ Q_+ - \chi_{BH} \biggl[
\sin \Theta \sin \Phi (m\omega)^{2/3} \nonumber \\ && \mbox{}
- {4 \over 3} \left( 1 + \cos^2
\Theta \right) \cos 2\Phi (m\omega) \biggr] \biggr\}
\end{eqnarray}
\begin{eqnarray}
h_\times = && {2\mu \over D} (m\omega)^{2/3} \biggl\{ Q_\times + \chi_{BH}
\biggl[
\sin \Theta \cos \Theta \cos \Phi (m\omega)^{2/3} \nonumber \\ && \mbox{}
+ {8 \over 3}  \cos
\Theta \sin 2\Phi (m\omega) \biggr] \biggr\}
\end{eqnarray}
where we have rewritten the equations using our notation.
\end{mathletters}

Combining Eqs. (\ref{circedot}) and (\ref{circr}) we find the energy loss to
be
\begin{eqnarray}
{dE \over dt} = && -{32 \over 5} \eta^2 (m\omega)^{10/3} \Biggl\{ 1 -
{1 \over 4} \sum_{i=1,2} \chi_i ({\bf \hat L_N \cdot \hat s_i})
\nonumber \\ && \mbox{} \times
\left( 11 {m_i^2 \over m^2} + 5 \eta \right) (m\omega) \Biggr\}
\label{tmedot}
\end{eqnarray}

In the test-mass limit that we are considering, $m_1 = \mu$, $m_2 = m$ (with
$\mu << m$), $\eta = \mu/m$, $\chi_1 = 0$, $\chi_2 = \chi_{BH}$, $\delta m=-m$,
${\bf \hat s_2} = {\bf \hat L_N}$, and
${\bf \Delta} = \chi_{BH}m^2{\bf \hat L_N}$.  Substituting these values into
Eq. (\ref{tmedot}) we see that our expression for the
energy loss agrees with that
of Poisson in the appropriate limit.

Combining Eqs. (\ref{circh}) and (\ref{circr}), and
taking the above test mass limit of the spin contributions yields
\begin{eqnarray}
h_c^{ij} &=& {2\mu \over D} (m\omega)^{2/3} \left\{ Q_c^{ij} + 2\chi_{BH}
(m\omega)^{2/3} n^{(i} ({\bf \hat L \times \hat N})^{j)} \right. \nonumber \\
&& \left. + {8 \over 3} \chi_{BH} (m\omega) \left[
n^in^j -  \lambda^i \lambda^j
\right] \right\}.
\end{eqnarray}
Substituting ${\bf \hat n} = \cos \Phi {\bf \hat x} + \sin \Phi {\bf \hat y}$
and ${\bf \hat \lambda} = - \sin \Phi {\bf \hat x} + \cos \Phi {\bf \hat y}$,
into the above expression and inserting the results into Eqs. (\ref{h+hx})
we obtain a result which matches that of Poisson.

\begin{figure}
\caption{Amplitude modulation of gravitational waveforms by spin-induced
orbital precession, plotted against time to coalescence.  System consists of
a nonspinning $1 M_\odot$ neutron star and a maximally spinning black
hole of $10 M_\odot$.  Spin and orbital angular momentum vectors are initially
misaligned by $11.3^\circ$.  Initial orbital inclination relative to
${\bf J}$ is $i$.  The angle $\gamma$ represents the orientation of the
detector relative to ${\bf J}$, with the detector located on the
x-axis (see Fig. \protect\ref{source_coordinates})
such that the source is directly overhead.
Curves show envelope of the quadrupole waveform for various detector
orientations (The curves for the cases $\gamma = 0$, $i/2$, and $i$ would
lie on top of one another, so the first two have been shifted upward for ease
of presentation).  Gravitational-wave frequency runs from 10 Hz on the
right to 300 Hz on the left.}
\label{amplitude_modulations}
\end{figure}

\begin{figure}
\caption{The source coordinate system.  The total
angular momentum ${\bf J}$
initially lies along the z-axis.  The detector is located in the x-z plane.
The spherical angles $(i,\alpha)$ define the direction of the Newtonian
angular momentum ${\bf L_N}$ which is perpendicular to the orbital plane.
In terms of celestial mechanics, the angle of ascending nodes is $\alpha
+ \pi/2$.}
\label{source_coordinates}
\end{figure}

\begin{figure}
\caption{A comparison of the magnitudes of the orbital angular momentum
and spin angular momentum as the binary inspirals.  (a) The equal
mass case (assuming the bodies are
maximally spinning).  (b) The case of a 10:1 mass ratio.}
\label{momentum_size}
\end{figure}

\begin{figure}
\caption{Gravitational wavefrom plotted against orbital phase for a
10:1.4 mass-ratio system.  The smaller body's spin is aligned with
the orbital angular momentum ${\bf L}$, while the larger body's spin
is tilted by an angle of $30^\circ$ with respect to ${\bf L}$.
Plotted is $(D/2\mu)h_+$ for an observer at $\Theta = 90^\circ$.
Plots begin at an orbital separation of $15m$ and terminate at $10m$.
(a) The total waveform.
(b) The quadrupole contribution to the waveform.
(c) The first higher-order post-Newtonian correction (${\cal O}(\epsilon^{1/2}
)$ beyond the quadrupole).
(d) The next post-Newtonian correction.
(e) The leading-order spin-orbit contribution to the waveform.
(f) The leading-order spin-spin contribution to the waveform
(note the different scale).
Notice the modulation due to the precession of the orbital plane.}
\label{gravitational_waveform}
\end{figure}

\begin{figure}
\caption{The wobble of the orbital plane during simple precession (in the
absence of gravitational-radiation damping).  Plotted are the $x$- and
$y$-components of the unit vector ${\bf \hat L_N}$ which defines the
orbital plane.  The total angular momentum ${\bf J}$ is directed
out of the page.  If the orbital plane were not wobbling as it precessed,
the plot would be a circle whose radius depends on the inclination of the
orbital plane with respect to ${\bf J}$.}
\label{wobble}
\end{figure}

\begin{figure}
\caption{The complicated precession of the orbital angular momentum ${\bf L}$
(in the absence of gravitational-radiation damping) for a 2:1 mass-ratio
system with equal spins which are initially aligned in the orbital plane.
Although the spins are equal, they precess at different rates so that at a
later time they almost cancel one another.  This leads to the tilting of
the orbital plane as it precesses about ${\bf J}$.
See Fig. \protect\ref{comp_prec_rad} for the effects
of gravitational radiation on the precession.}
\label{comp_prec_norad}
\end{figure}

\begin{figure}
\caption{The energy lost due to gravitational radiation plotted
against orbital phase for a
10:1.4 mass-ratio system (with spins initially aligned as in
Fig. \protect\ref{gravitational_waveform}).
Plots begin at an orbital separation of $15m$ and terminate at $10m$.
Plotted are the lowest-order Newtonian (N), post-Newtonian (PN),
spin-orbit (SO), and spin-spin (SS) contributions to the energy loss.}
\label{luminosity}
\end{figure}

\begin{figure}
\caption{The linear momentum ejected due to gravitational radiation plotted
against orbital phase for a
10:1.4 mass-ratio system (with spins initially aligned as in
Fig. \protect\ref{gravitational_waveform}).
Plots begin at an orbital separation of $15m$ and terminate at $10m$.}
\label{momentum_ejection}
\end{figure}

\begin{figure}
\caption{The simple precession of the orbital angular momentum ${\bf L}$
in the presence of gravitational radiation damping
for an inspiraling system with a 2:1 mass-ratio with only the larger body
spinning
(The spin is perpendicular to ${\bf L}$).
Notice that the inclination
of the orbital plane with respect to ${\bf J_o}$
increases as the binary inspirals
($i = \sin^{-1}[(\hat L_x^2 + \hat L_y^2)^{1/2}]$).}
\label{simple_precession}
\end{figure}

\begin{figure}
\caption{The complicated precession of the orbital angular momentum ${\bf L}$
in the presence of gravitational radiation damping
for an inspiraling system with a 2:1 mass-ratio and equal spins
which are initially aligned, and perpendicular to ${\bf L}$.
Notice that the maximum inclination
of the orbital plane increases as the binary inspirals.}
\label{comp_prec_rad}
\end{figure}

\begin{figure}
\caption{The detector coordinate system.  The detectors arms lie along the
$\bar x$- and $\bar y$-axes.  The angles $(\theta,\phi)$ define the location
of the source, and the polarization angle $\psi$ describes the orientation
of the polarization axes with respect to the projection of the detectors arms
on the sky.}
\label{detector_coordinates}
\end{figure}

\begin{figure}
\caption{The region of validity for the small inclination approximation for
an equal mass system.
The figure plots the minimum separation $r/m$ for which the precession
angle $i$ is guaranteed to be less than $i_{max}$ as a function of
the angle between ${\bf L}$ and ${\bf S}$ for several values of the
average spin parameter $\chi$ of the two bodies.  For most neutron star
models $\chi \leq 0.7$. Note that at large angles, ``transitional
precession" may occur.}
\label{small_inc}
\end{figure}

\begin{figure}
\caption{The amplitude modulations of the higher-order terms of the
gravitational waveform.  $A_1$ and $A_3$ are of (post)$^{1/2}$-Newtonian order
beyond the quadrupole term, while $A_4$ is (post)$^1$-Newtonian order.  The
quadrupole envelope increase from $0.06$ to $0.36$ during the portion of the
inspiral which is plotted.  Note that since the different terms have different
dependences on the orbital phase,  the overall waveform envelope will not
be the sum of the individual envelopes.  The binary system is the same as
in Fig. \protect\ref{amplitude_modulations}, but for an observer at
$\Theta=\pi/4$,
with $\gamma = 0$.}
\label{high-order_modulations}
\end{figure}

\begin{figure}
\caption{The rate of precession for the two-spin case and the
corresponding one-spin case
for a 10:1 mass-ratio system with the total spin ${\bf S}$ at
an angle of $\pi/4$ with the orbital angular momentum ${\bf L}$.  In the
two-spin case, the smaller body's spin is aligned with ${\bf L}$.
In both cases the bodies are maximally spinning.}
\label{large_mass-ratio}
\end{figure}

\begin{figure}
\caption{An equal mass case in which the spins of the bodies are equal in
magnitude
($\chi_1 = \chi_2 = \protect\sqrt{2}/2$),
with one spin aligned with ${\bf L}$ and the other perpendicular to it.
(a) The modulation of the quadrupole waveform envelope.
(b) The components of the total angular momentum ${\bf J}$.
Notice that ${\bf J}$ remains relatively fixed in direction
as the binary inspirals due to gravitational radiation emission.
Also plotted is the angle between ${\bf L}$ and ${\bf S}$, which oscillates
about a relatively constant value.
(c) The components of ${\bf S_1}$.
(d) The components of ${\bf S_2}$.
Notice that while the individual spins undergo substantial
precessions, the total spin precesses in a fairly simple manner.
This causes the two-spin case to be similar to the corresponding
one-spin case.}
\label{equal_mass}
\end{figure}

\begin{figure}
\caption{The modulation of the quadrupole waveform envelope for a
2:1 mass-ratio system with the total spin ${\bf S}$ initially
perpendicular to ${\bf L}$.  (a) One spinning body.
(b) Two spinning bodies with spins equal such that the smaller
body's spin is maximal ($\chi = 1$).
Note the substantial differences in the modulations between the
two cases.  This is a result of the complicated precession when both
bodies are spinning.  See Figs. \protect\ref{simple_precession} and
\protect\ref{comp_prec_rad} for a depiction of the
precession of the orbital angular momentum for these cases.}
\label{wild_precession}
\end{figure}

\mediumtext
\begin{table}
\caption{The contributions to the number of orbits for an inspiralling
binary system which is in a LIGO-type detector's bandwidth, from
quadrupole (Quad), post-Newtonian (PN), tail, spin-orbit (SO), and
spin-spin (SS) terms.  Masses are
in units of $M_\odot$.  In the first table, the observed inspiral
begins when the gravitational
wave frequency
enters the bandwidth at the seismic cutoff around 10 Hz, and
is cut off when either the quadrupole
radiation of the binary has left the detector's bandwidth (at around
1000 Hz), or
when the binary reaches its innermost stable circular orbit at which
point the binary will plunge to a single object in
one or two orbits.  In the second table, we list the contribution to the
number of orbits in the narrower bandwidth of 40 Hz to 100 Hz.
The spin contributions shown are the maximum contributions
assuming the bodies are maximally spinning ($\chi = \chi_{max}$).
The magnitude and sign of the spin contributions depend
on the specific orientations of the spins with respect to the orbital plane.}
\label{phasetable}
\begin{tabular}{dddddddddd}
$m_1$&$m_2$&$r_{10 Hz}/m$&$f_{cut} (Hz)$&$r_{cut}/m$&Quad&PN&Tail&SO&SS\\
\tableline
1.4&1.4&174&1000&7.3&8015&219&-104&$\pm$44&$\pm$1.2 \\
0.9&1.8&178&1000&7.5&9581&249&-119&$\pm$53&$\pm$1.2 \\
0.5&2.0&187&1000&7.5&15128&351&-175&$\pm$82&$\pm$1.3 \\
1.4&10&68&345&5.6&1787&106&-90&$\pm$63&$\pm$1.0 \\
1.4&100&15&42&5.1&337&66&-112&$\pm$83&$\pm$0.3 \\
10&10&46&183&6.0&300&29&-25&$\pm$15&$\pm$1.0 \\
5&10&56&245&6.0&547&43&-35&$\pm$22&$\pm$1.2 \\
\end{tabular}
\begin{tabular}{ddddddddd}
$m_1$&$m_2$&$r_{40 Hz}/m$&$r_{100 Hz}/m$&Quad&PN&Tail&SO&SS\\
\tableline
1.4&1.4&68&37&623&33&-19&$\pm$12&$\pm$0.5 \\
0.9&1.8&70&38&745&37&-22&$\pm$14&$\pm$0.5\\
0.5&2.0&74&39&1175&531&-33&$\pm$22&$\pm$0.6 \\
1.4&10&26&14&139&16&-18&$\pm$13&$\pm$0.3 \\
10&10&18&9.3&24&4.7&-5.3&$\pm$3.3&$\pm$0.3 \\
5&10&22&11&43&6.8&-7.3&$\pm$4.6&$\pm$0.3 \\
\end{tabular}
\end{table}

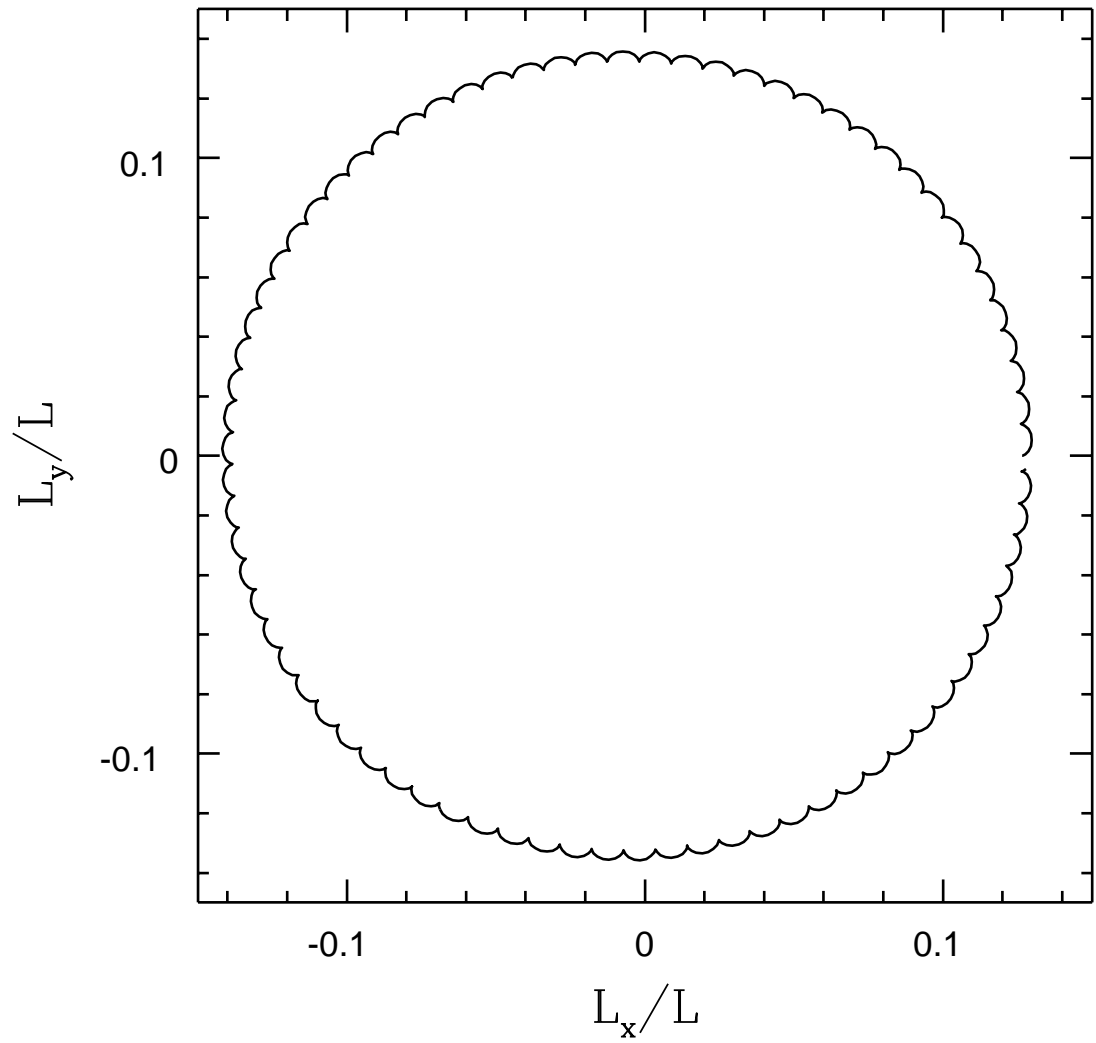
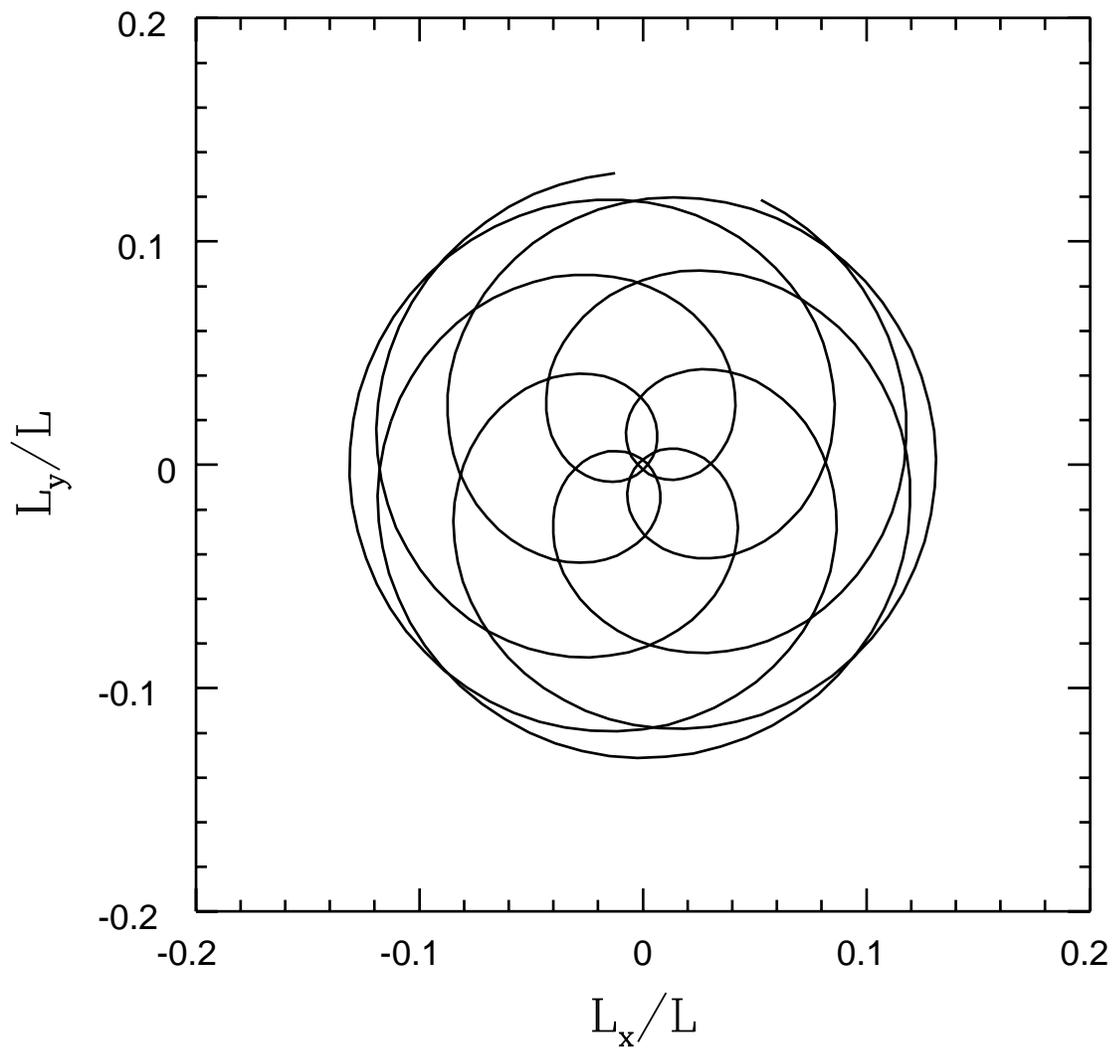
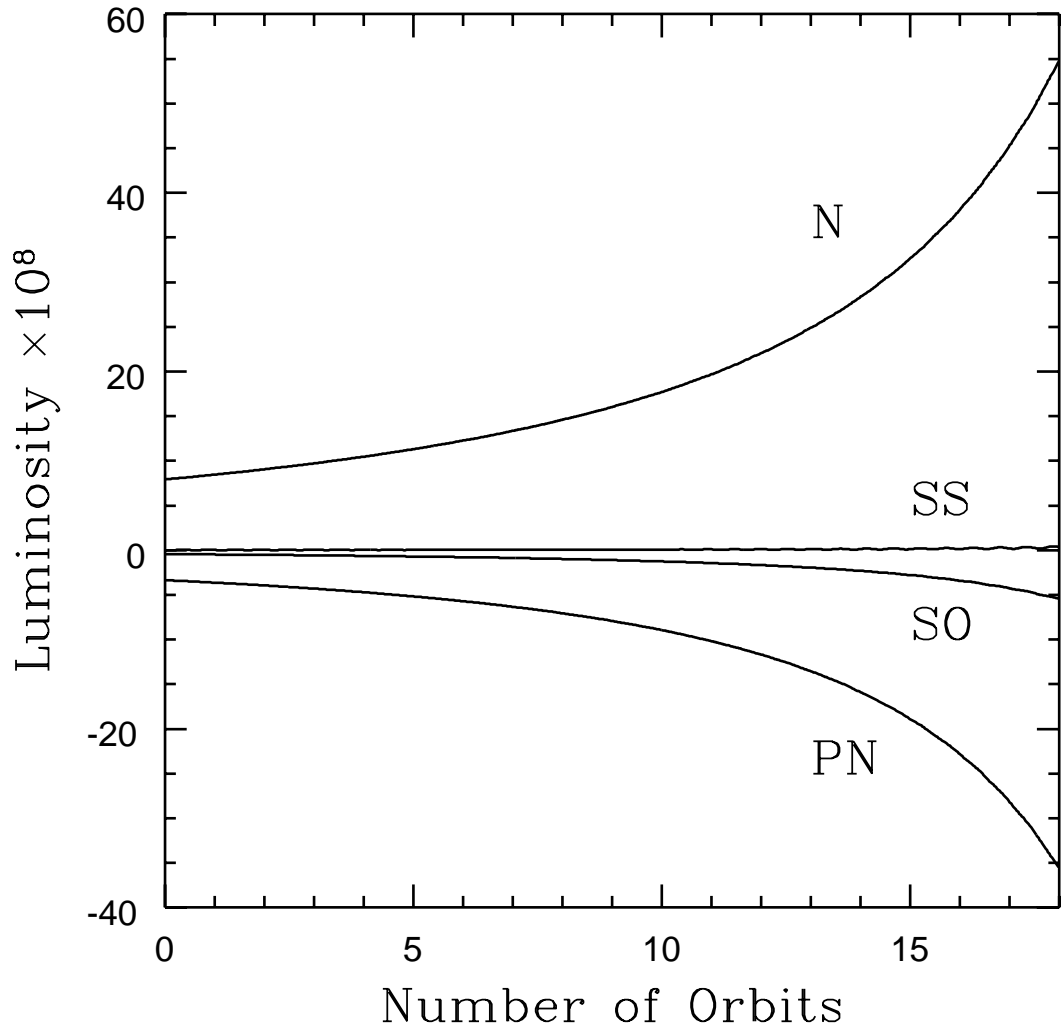
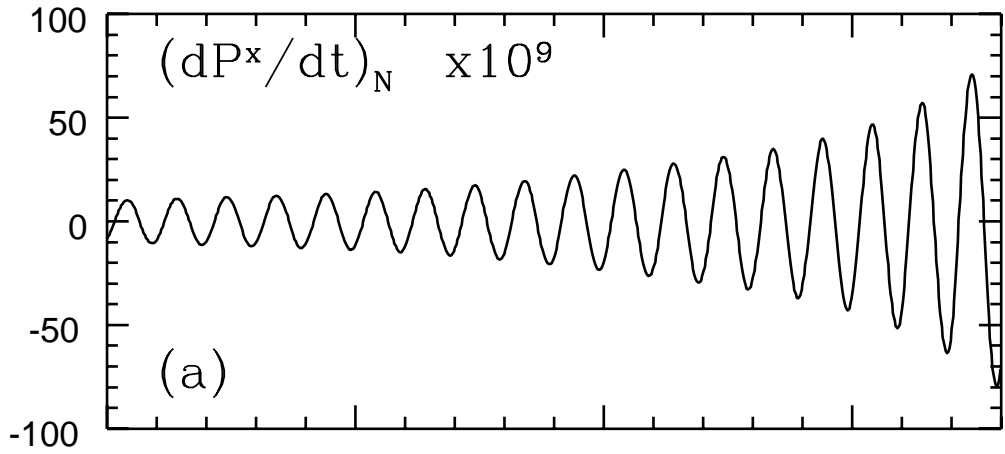
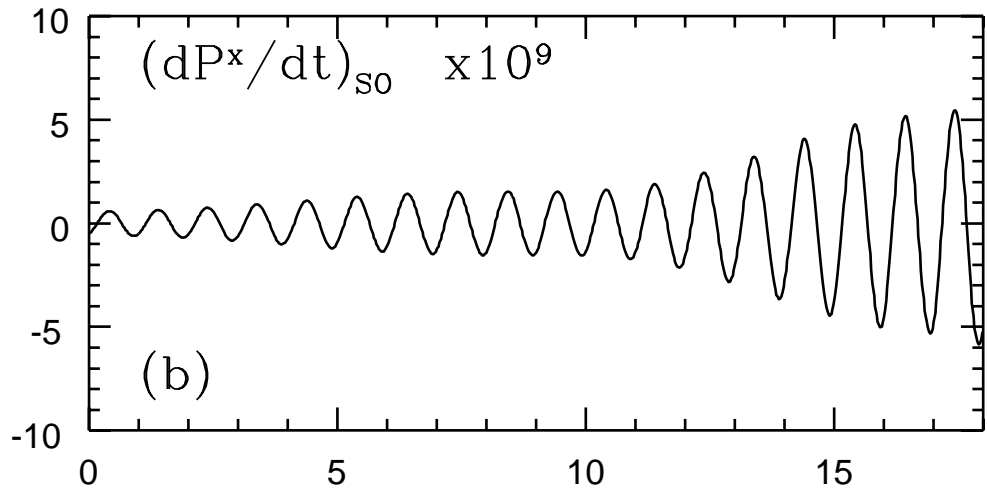
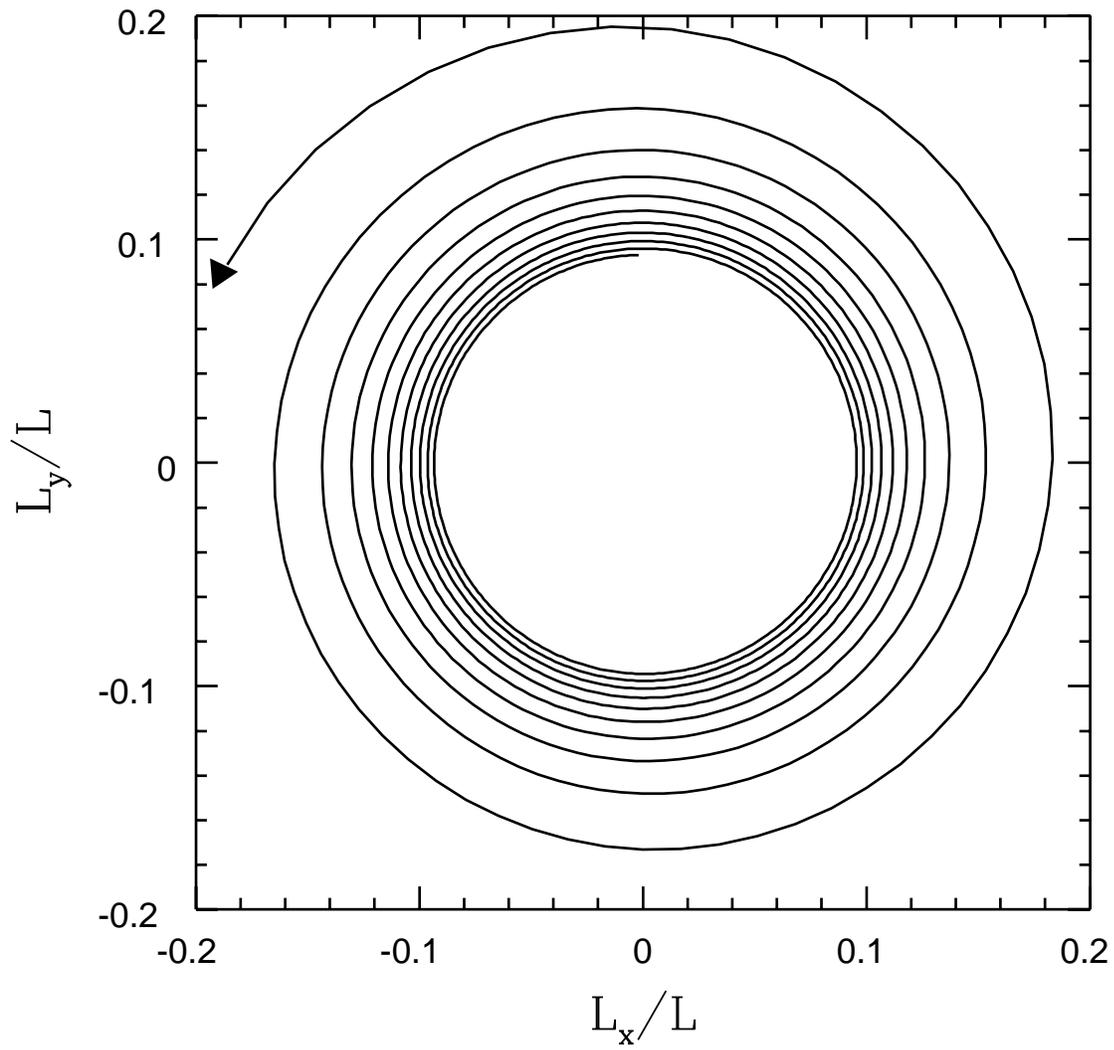
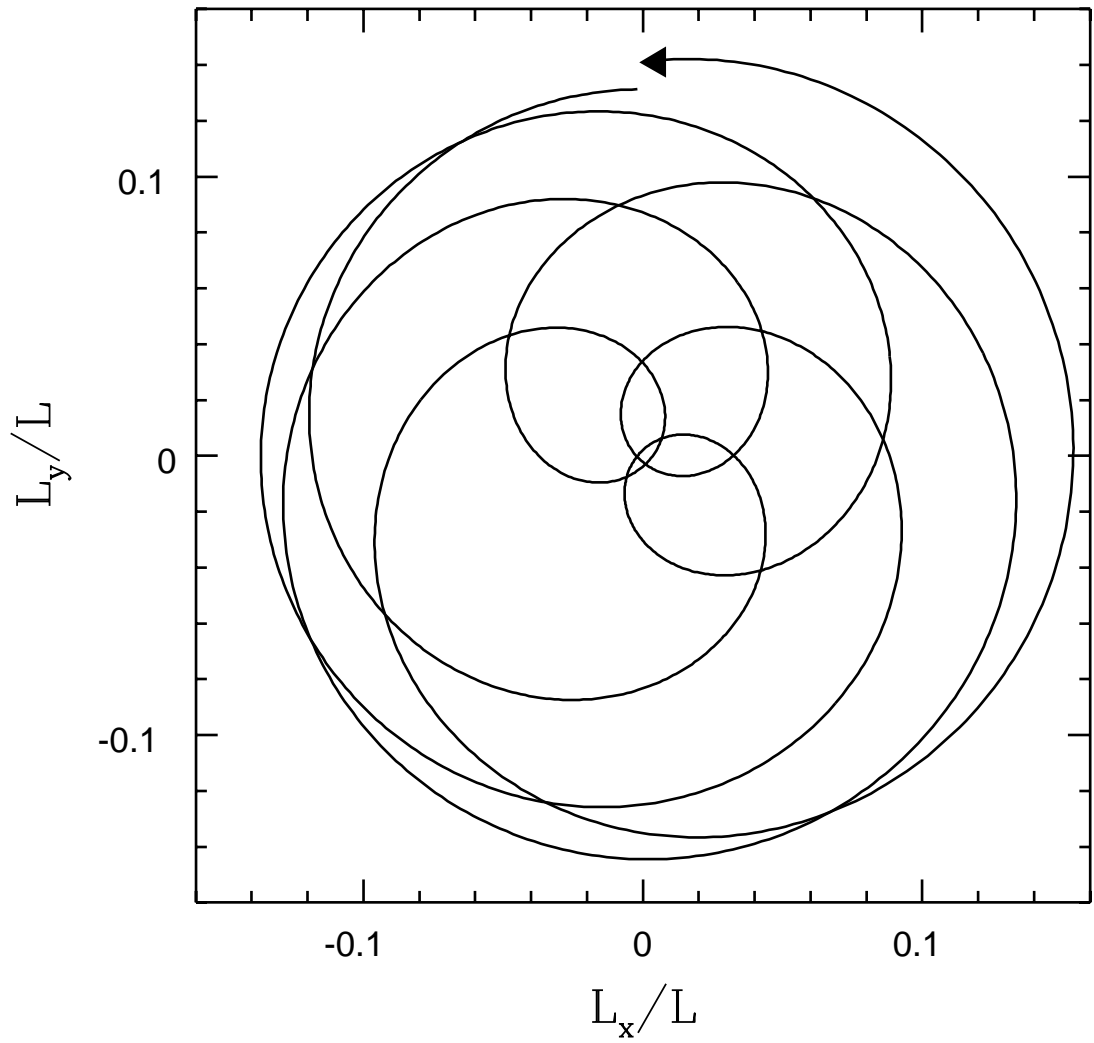
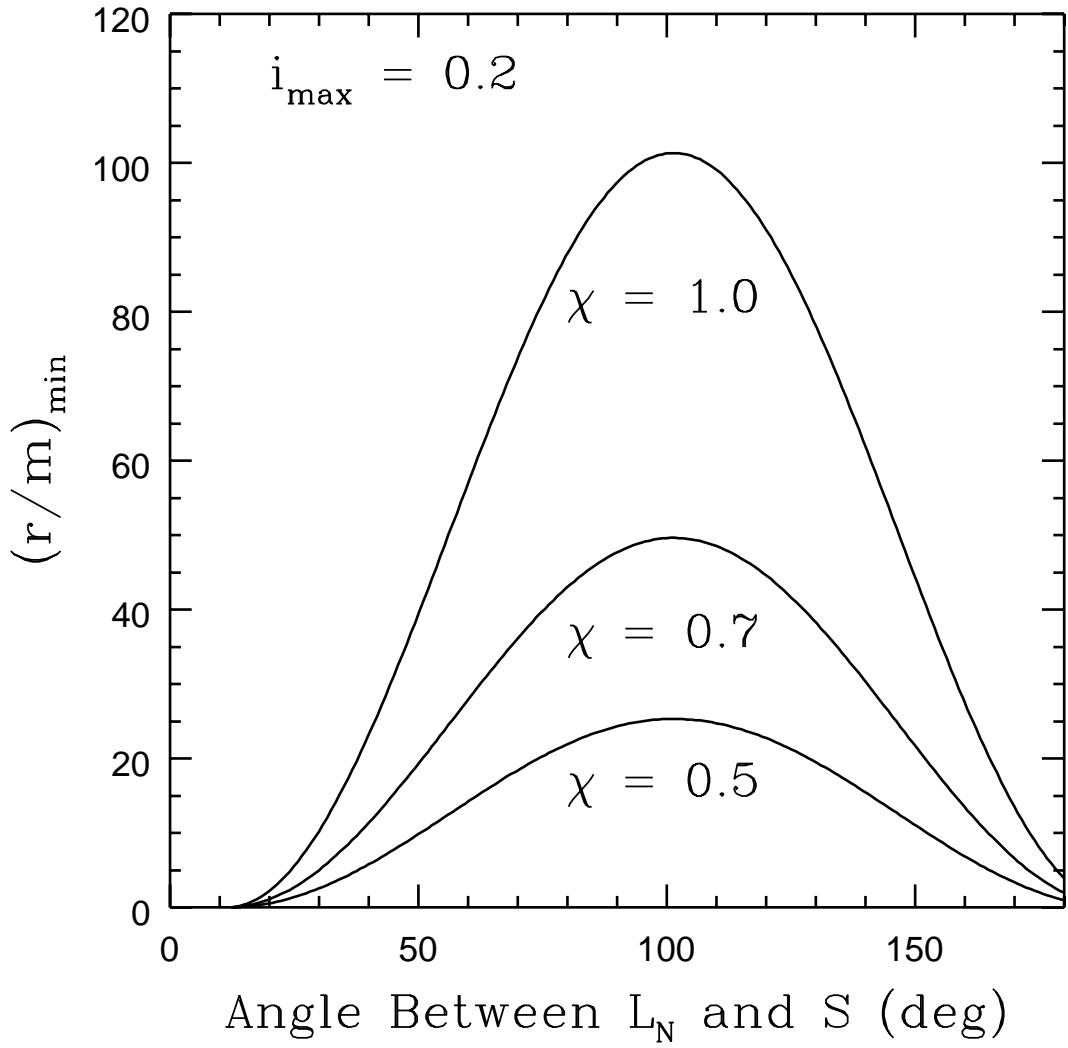
\end{document}